\documentclass[
  aps,
  reprint,
  onecolumn,
  superscriptaddress,
  amsmath,
  amssymb,
  nofootinbib
]{revtex4-2}

\usepackage{graphicx}
\usepackage{dcolumn}
\usepackage{bm}

\makeatletter

\newcommand{\authornotes}[1]{%
  \gdef\revtex@authornotes{#1}%
}

\def\doauthor#1#2#3{%
  \global\let\revtex@authornotes\@empty
  \ignorespaces#1\unskip\@listcomma
  \begingroup
    #3%
    \revtex@authornotes
  \@if@empty{#2}
    {\endgroup{}{}}
    {\endgroup{\comma@space}{}\frontmatter@footnote{#2}}%
  \space\@listand
}

\makeatother

\begin{document}

\title{Resolving Scale-Dependent Diffusivity in the Brain Extracellular Space}

\author{Quentin Gresil%
  \authornotes{\textsuperscript{,\textdagger,*}}}
\affiliation{Laboratoire Photonique Num\'erique et Nanosciences,
Universit\'e de Bordeaux, 33400 Talence, France}
\affiliation{LP2N, Institut d'Optique Graduate School,
CNRS UMR 5298, 33400 Talence, France}

\author{Ivo Calaresu}
\affiliation{Interdisciplinary Institute for Neuroscience, CNRS, Univ. Bordeaux, 33076 Bordeaux, France}

\author{Finn L. Sebastian}
\affiliation{Institute for Physical Chemistry, Universität Heidelberg, Heidelberg, Germany}

\author{Benjamin Flavel}
\affiliation{Institute of Nanotechnology, Karlsruhe Institute of Technology, 76131 Karlsruhe, Germany}

\author{Jana Zaumseil}
\affiliation{Institute for Physical Chemistry, Universität Heidelberg, Heidelberg, Germany}

\author{Laurent Groc}
\affiliation{Interdisciplinary Institute for Neuroscience, CNRS UMR 5297, Univ. Bordeaux, 33077 Bordeaux, France}

\author{Laurent Cognet%
  \authornotes{\textsuperscript{,*}}}
\affiliation{Laboratoire Photonique Num\'erique et Nanosciences,
Universit\'e de Bordeaux, 33400 Talence, France}
\affiliation{LP2N, Institut d'Optique Graduate School,
CNRS UMR 5298, 33400 Talence, France}

\date{\today}

\begin{abstract}

Transport through the brain extracellular space has traditionally been
summarized by effective diffusion coefficients measured over specific
observation ranges. Whether local mobility remains predictive as the same
molecule explores larger distances remains unresolved. Here, we track
individual ultrashort carbon nanotubes in three dimensions within living
hippocampal tissue, following their motion from nanometre to micrometre scales.
Using freely diffusing nanotubes in water as an experimental reference, we
resolve trajectory-specific crossover lengths beyond which instantaneous
diffusivity decreases, with slice-level medians of
$0.67~\mu\mathrm{m}$ in the pyramidal layer and $0.52~\mu\mathrm{m}$ in the
radiatum. The pyramidal layer combines higher short-time diffusivity with an
approximately twofold larger post-crossover restriction exponent than the
radiatum ($\vartheta=0.77$ versus $0.38$). This decoupling of local mobility
from larger-scale exploration is incompatible with a scale-independent
rescaling of transport, showing that effective extracellular diffusivity must
be interpreted relative to the scale of exploration.

\end{abstract}

\begingroup
\renewcommand{\thefootnote}{\fnsymbol{footnote}}
\footnotetext[2]{Present address: Laboratory of Nanoscale Biology (LBEN), Institute of Bioengineering, School of Engineering, Swiss Federal
Institute of Technology Lausanne (EPFL), Lausanne, Switzerland.}
\footnotetext[1]{Correspondence and requests for materials should be
addressed to: Q.G. (\texttt{quentin.gresil@epfl.ch}) or
L.C. (\texttt{laurent.cognet@u-bordeaux.fr}).}
\endgroup

\maketitle


Molecular transport through the brain extracellular space (ECS) supports
neurotransmitter signaling, metabolite exchange, and the delivery of
therapeutic agents
\cite{nicholsonExtracellularSpaceStructure1998,
sykovaDiffusionBrainExtracellular2008a}.
It also contributes to the clearance of metabolites and other solutes from
brain tissue
\cite{xieSleepDrivesMetabolite2013}.
These processes depend on how molecules explore the interconnected spaces
between neurons, glia, and blood vessels, which together occupy approximately
20\% of brain tissue
\cite{sykovaDiffusionBrainExtracellular2008a,
nicholsonBrainExtracellularSpace2017a,
tonnesenLocalDiffusionExtracellular2023}.
Determining how this geometrically complex network regulates molecular motion
is therefore central to understanding both brain physiology and neurological
disease.

Extracellular transport is commonly described by effective parameters such as
diffusivity, tortuosity, and extracellular volume fraction. These quantities
are obtained using methods with distinct molecular sensitivities and
observation ranges, including diffusion MRI, real-time iontophoresis, and
single-particle tracking
\cite{novikovQuantifyingBrainMicrostructure2019,
nicholsonExtracellularSpaceStructure1998,
godinSingleNanotubeTrackingReveals2017a}.
Their interpretation therefore depends on the spatial and temporal scales
sampled by each method. This dependence is explicitly incorporated in
diffusion MRI, where changes in apparent diffusivity with diffusion time are
used to probe tissue microstructure
\cite{novikovQuantifyingBrainMicrostructure2019}.
Yet, at the level of individual extracellular probes, transport has generally
been characterized using parameters evaluated over selected ranges of lag
times
\cite{godinSingleNanotubeTrackingReveals2017a,
soriaSynucleinopathyAltersNanoscale2020,
grassiNanoscaleFunctionalHeterogeneity2023}.
How local mobility evolves as the same probe explores progressively larger
distances therefore remains unknown.

This distinction becomes important when transport parameters obtained by
different methods are used to infer changes in extracellular structure or
physiology. Disruption of extracellular matrix components, for example, was
reported to increase extracellular volume fraction measured by iontophoresis
while decreasing tissue water diffusivity measured by diffusion MRI
\cite{sykovaDisruptionExtracellularMatrix2025}.
Because the two measurements probe different molecules, compartments, and
scales, their opposing responses cannot be attributed unambiguously to either
methodological specificity or scale-dependent transport
\cite{sykovaDiffusionBrainExtracellular2008a,
nicholsonBrainExtracellularSpace2017a}.
Distinguishing these possibilities requires measuring transport within the
same trajectories, from local displacements to larger-scale exploration of
the extracellular network.

Here, we combine ultrashort carbon nanotube probes
\cite{setteleEasyAccessBright2024,
nandiUltrashortCarbonNanotubes2025b}
with three-dimensional single-particle tracking in living hippocampal tissue
\cite{pavaniThreedimensionalSinglemoleculeFluorescence2009,
vondiezmannThreeDimensionalLocalizationSingle2017}.
This experimental capability allows us to follow the same extracellular
probes within individual trajectories, from nanometre-scale displacements to
micrometre-scale exploration, and to determine how their apparent mobility
evolves with scale. Freely diffusing nanotubes in water provide an
experimental reference for identifying departures from Brownian MSD scaling
within individual trajectories. We then determine when these departures
become detectable, how diffusivity evolves thereafter, and whether these
properties differ between hippocampal regions. This approach introduces a
trajectory-resolved framework for distinguishing local mobility from its
subsequent evolution with scale.

\section*{Results}

\noindent\textbf{Extracellular trajectories progressively depart from Brownian
scaling.}
We measured brain extracellular transport in organotypic slices (see Methods) by
tracking individual ultrashort carbon nanotubes (uCCNTs) using near-infrared
three-dimensional single-particle microscopy
\cite{pavaniThreedimensionalSinglemoleculeFluorescence2009,
vondiezmannThreeDimensionalLocalizationSingle2017}
with a localization precision of 10--20 nm (Fig.~S1). uCCNTs have
characteristic lengths of approximately 50 nm
\cite{setteleEasyAccessBright2024,nandiUltrashortCarbonNanotubes2025b},
making them an order of magnitude shorter than nanotube probes used previously
\cite{godinSingleNanotubeTrackingReveals2017a,
soriaSynucleinopathyAltersNanoscale2020}.
Their dimensions overlap reported extracellular space widths of approximately
50--500 nm
\cite{korogodUltrastructuralAnalysisAdult2015,
tonnesenSuperResolutionImagingExtracellular2018,
soriaSynucleinopathyAltersNanoscale2020}.
Their short length is therefore expected to reduce, although not eliminate,
probe-size-dependent steric hindrance relative to longer nanotube probes
\cite{thorneInVivoDiffusionAnalysis2006,
nicholsonBrainExtracellularSpace2017a}.

\begin{figure*}[htbp!]
\centering
\includegraphics[scale=1]{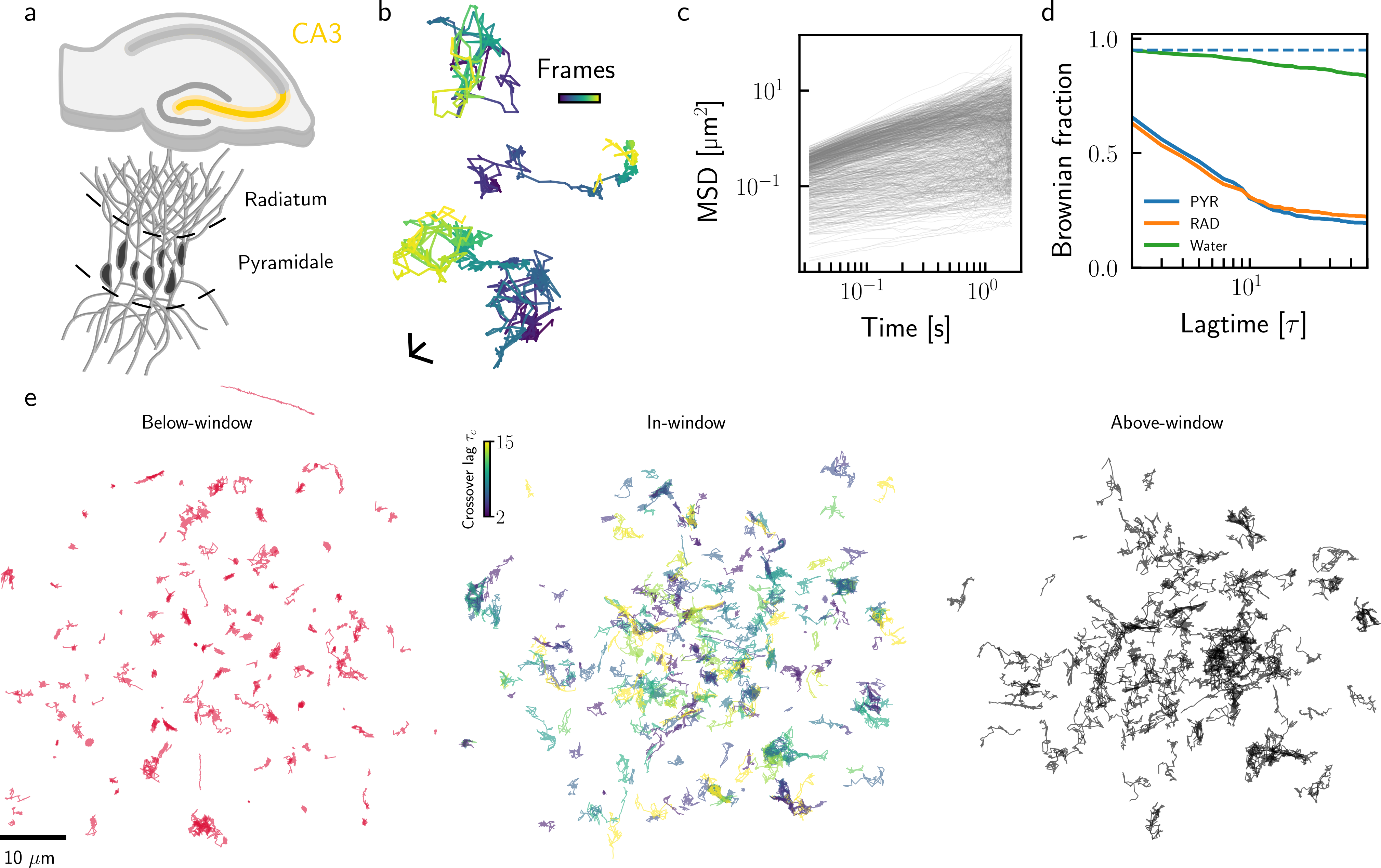}
\caption{\label{fig:brownian_departure}
\textbf{Trajectory-resolved detection of departures from Brownian diffusion
scaling in the hippocampal extracellular space.}
(a) Schematic of the hippocampal CA3 region and the cellular organization of
the stratum radiatum and stratum pyramidale, where extracellular nanotube
motion was recorded.
(b) Representative three-dimensional trajectories of individual ultrashort
carbon nanotubes, color-coded by acquisition frame.
(c) Broad diversity of time-averaged mean squared displacement curves from
individual trajectories.
(d) Fraction of trajectories remaining compatible with Brownian mean squared
displacement scaling as the fitted lag-time window is extended for the
pyramidal layer (PYR), radiatum (RAD), and freely diffusing nanotubes in
water. Compatibility was determined using the distribution of fitting errors
measured in water. The dashed horizontal line indicates the nominal 95\%
acceptance level.
(e) Spatial projections of trajectories grouped according to whether
departure from Brownian scaling was detected at the earliest tested window,
at a resolved later crossover, or not detected within the observation
window. \emph{In-window} trajectories are color-coded by their crossover lag,
$\tau_c/\Delta t$. Scale bar, $10~\mu\mathrm{m}$. The acquisition interval was
$\Delta t=33~\mathrm{ms}$.
}
\end{figure*}

We focused on two layers of hippocampal CA3 with contrasting cellular
organization (Fig.~\ref{fig:brownian_departure}a). The pyramidal layer (PYR)
is dominated by densely packed neuronal somata, whereas the stratum radiatum
(RAD) consists primarily of neuropil, including dendrites, axons, synapses,
glial processes, and blood vessels
\cite{gahwilerOrganotypicCulturesNeural1988,
grassiNanoscaleFunctionalHeterogeneity2023}.
Individual uCCNT trajectories exhibited substantial variation in spatial
extent and morphology (Fig.~\ref{fig:brownian_departure}b). This diversity
was also apparent in their time-averaged mean squared displacements, which
spanned several orders of magnitude and differed markedly in their evolution
with lag time (Fig.~\ref{fig:brownian_departure}c).

A trajectory described by a single, scale-independent diffusion coefficient must exhibit a mean squared displacement proportional to lag time throughout the observation window. We
therefore tested whether this proportionality was progressively lost as
larger lag times were included. Freely diffusing uCCNTs in water
($n=591$ trajectories) provided the experimental reference for this test.
For each trajectory, a Brownian model was fitted over progressively increasing
lag-time windows. The corresponding fitting error was compared with the
distribution measured from water trajectories over the same window. A
trajectory was considered compatible with Brownian scaling when its error
remained within the water-calibrated acceptance range. The window-dependent
fitting errors and empirical water thresholds are shown in Fig.~S2. Once the
acceptance criterion was exceeded, the classification was treated as
irreversible, preventing fluctuations at poorly sampled long lag times from
generating repeated transitions between accepted and rejected states.

The fraction of trajectories being compatible with Brownian scaling
decreased as the fitted lag-time window increased in both hippocampal regions,
whereas most water trajectories remained compatible over the same interval
(Fig.~\ref{fig:brownian_departure}d). The lag time at which departure was
detected varied substantially among extracellular trajectories.

We therefore classified trajectories into three operational, window-relative
categories according to the Brownian compatibility test
(Fig.~\ref{fig:brownian_departure}e). \emph{Below-window} trajectories were
already incompatible with Brownian scaling at the earliest tested fitting
window. \emph{In-window} trajectories were initially compatible with Brownian
scaling but subsequently exhibited a sustained departure at a resolved
crossover lag, $\tau_c$. \emph{Above-window} trajectories remained compatible
with Brownian scaling throughout the measured interval. These terms define
observational categories relative to the experimental window rather than
assuming a priori distinct transport mechanisms. Among the 721 PYR
trajectories, 246 (34.1\%) were classified as below-window, 335 (46.5\%) as
in-window, and 140 (19.4\%) as above-window. Among the 958 RAD trajectories,
the corresponding numbers were 354 (37.0\%), 392 (40.9\%), and 212 (22.1\%),
respectively (Supplementary Table~S1). Thus, 80.6\% of PYR trajectories and
77.9\% of RAD trajectories were either already incompatible with Brownian MSD
scaling at the earliest tested window or became incompatible within the
measured lag-time range.

These categories describe the observed mean squared displacement scaling
within the experimental window and provide the framework for the
trajectory-resolved analyses that follow.\\

\begin{figure}[htbp!]
\centering
\includegraphics[scale=1]{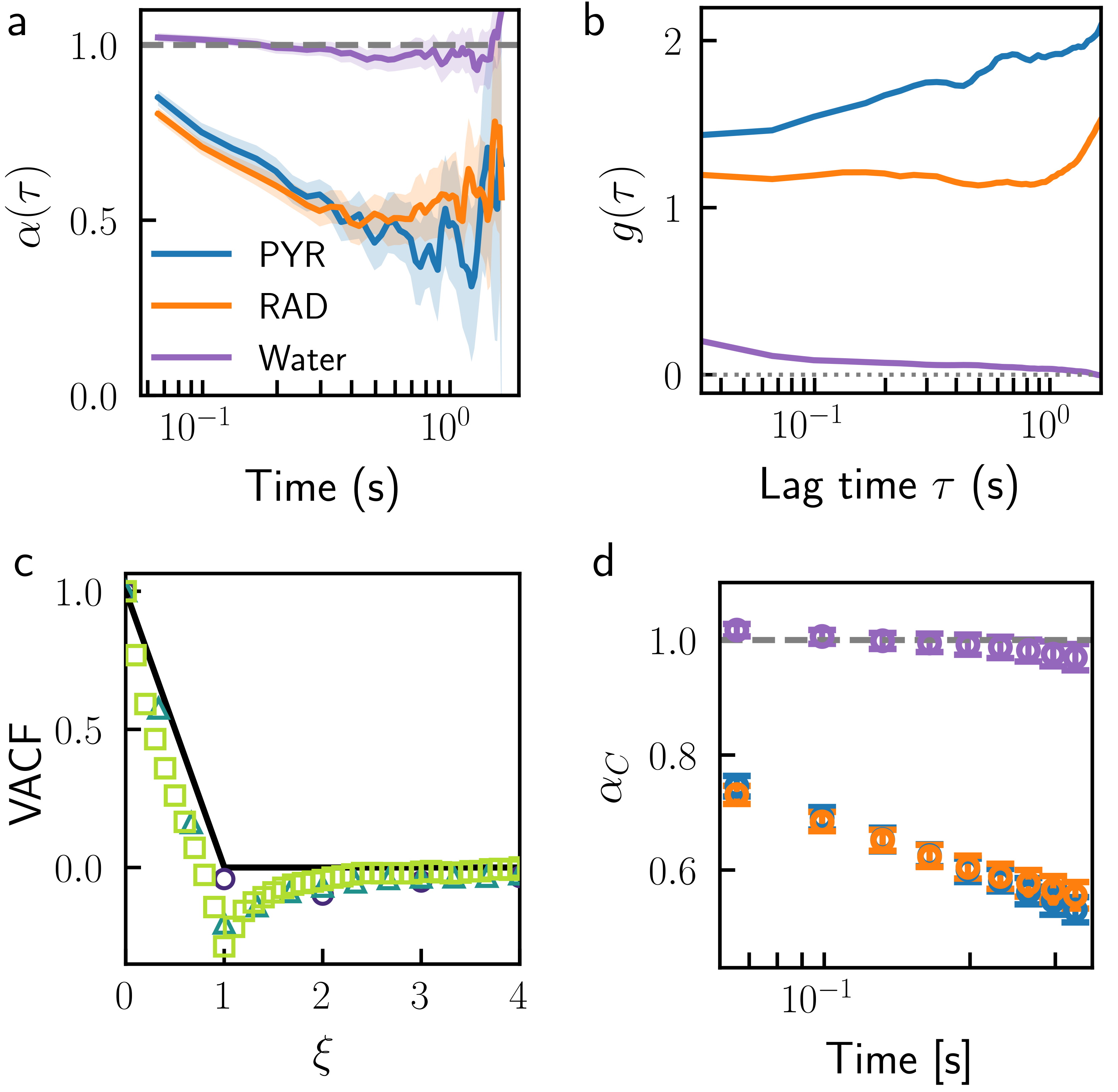}
\caption{\label{fig:resolved_departure_dynamics}
\textbf{\emph{In-window} trajectories develop subdiffusive, non-Gaussian,
and anticorrelated dynamics.}
(a) Local mean squared displacement scaling exponent, $\alpha(\tau)$, for
\emph{in-window} trajectories in PYR and RAD, with freely diffusing uCCNTs in
water shown as an experimental reference. The dashed line indicates the
Brownian value, $\alpha=1$, and shaded regions show 95\% trajectory-bootstrap
confidence intervals.
(b) Non-Gaussian parameter, $g(\tau)$, for the same populations. The dotted
line indicates the Gaussian expectation, $g=0$.
(c) Normalized velocity autocorrelation function (VACF) of PYR trajectories
for displacement intervals of 1, 3, and 10 frames, plotted against the
normalized separation $\xi$. The black line shows the expectation for
overlapping increments of Brownian motion. Negative correlations beyond the
overlap interval indicate preferential reversal of successive displacements.
(d) Effective increment-correlation exponent, $\alpha_C$, obtained by fitting
the fractional-Brownian increment-correlation form to the VACF
\cite{rehfeldtRandomWalkerToolbox2023}. The dashed line indicates the Brownian
expectation, $\alpha_C=1$, and error bars show 95\% trajectory-bootstrap
confidence intervals.
}
\end{figure}

\pagebreak

\noindent\textbf{\emph{In-window} trajectories reveal scale-dependent
transport hindrance and antipersistent motion.}
We began with trajectories whose crossover occurred within the experimental
window, because they provide a direct view of how extracellular transport
changes after an initially Brownian-compatible regime. Their local scaling
exponent,
$
\alpha(\tau)
=
\frac{\mathrm{d}\log\langle\Delta r^2(\tau)\rangle}
{\mathrm{d}\log\tau},
$
decreased below unity with increasing lag time in both PYR and RAD, whereas
water remained close to the Brownian expectation
(Fig.~\ref{fig:resolved_departure_dynamics}a). Transport therefore became
progressively more hindered as larger temporal and spatial scales were
explored, rather than being governed by a reduced but constant diffusion
coefficient
\cite{hoflingAnomalousTransportComplex2013,
rehfeldtRandomWalkerToolbox2023}.

We next asked whether this scale-dependent hindrance was accompanied by
directional memory. The velocity autocorrelation function developed a
pronounced negative dip, indicating preferential reversal of successive
displacements and therefore antipersistent motion
\cite{rehfeldtRandomWalkerToolbox2023}
(Fig.~\ref{fig:resolved_departure_dynamics}c). We summarized the scale
dependence of this behavior using the effective increment-correlation
exponent, $\alpha_C$. Its deviation from the Brownian value increased with
displacement interval in PYR and RAD, whereas water remained close to unity
(Fig.~\ref{fig:resolved_departure_dynamics}d). Here, $\alpha_C$ is used as a
descriptor of the observed correlation structure and does not identify
fractional Brownian motion as its underlying mechanism.

This hindrance was accompanied by increasingly non-Gaussian displacement
statistics. Mean squared displacement scaling and displacement Gaussianity
provide complementary, rather than equivalent, descriptions of stochastic
motion
\cite{wangAnomalousYetBrownian2009,
rehfeldtRandomWalkerToolbox2023}.
The non-Gaussian parameter, $g(\tau)$, remained positive in both hippocampal
regions and increased with lag time, particularly in PYR, while approaching
zero in water (Fig.~\ref{fig:resolved_departure_dynamics}b). Departure from
linear mean squared displacement growth was therefore accompanied by a change
in the displacement distribution.

The progressive emergence of transport hindrance was thus detected through
three complementary properties of the same trajectories: sublinear mean
squared displacement growth, non-Gaussian displacement amplitudes, and
anticorrelated increments. Their convergence shows that this scale dependence
affected both the magnitude and directional organization of molecular
displacements, rather than producing a uniform reduction of a Brownian
diffusion coefficient. We next quantified it through the initial diffusivity,
$D_0$, the crossover time, $\tau_c$, its associated exploration length,
$\ell_0$, and the post-crossover restriction exponent, $\vartheta$.\\

\begin{figure*}[htbp!]
\centering
\includegraphics[scale=1]{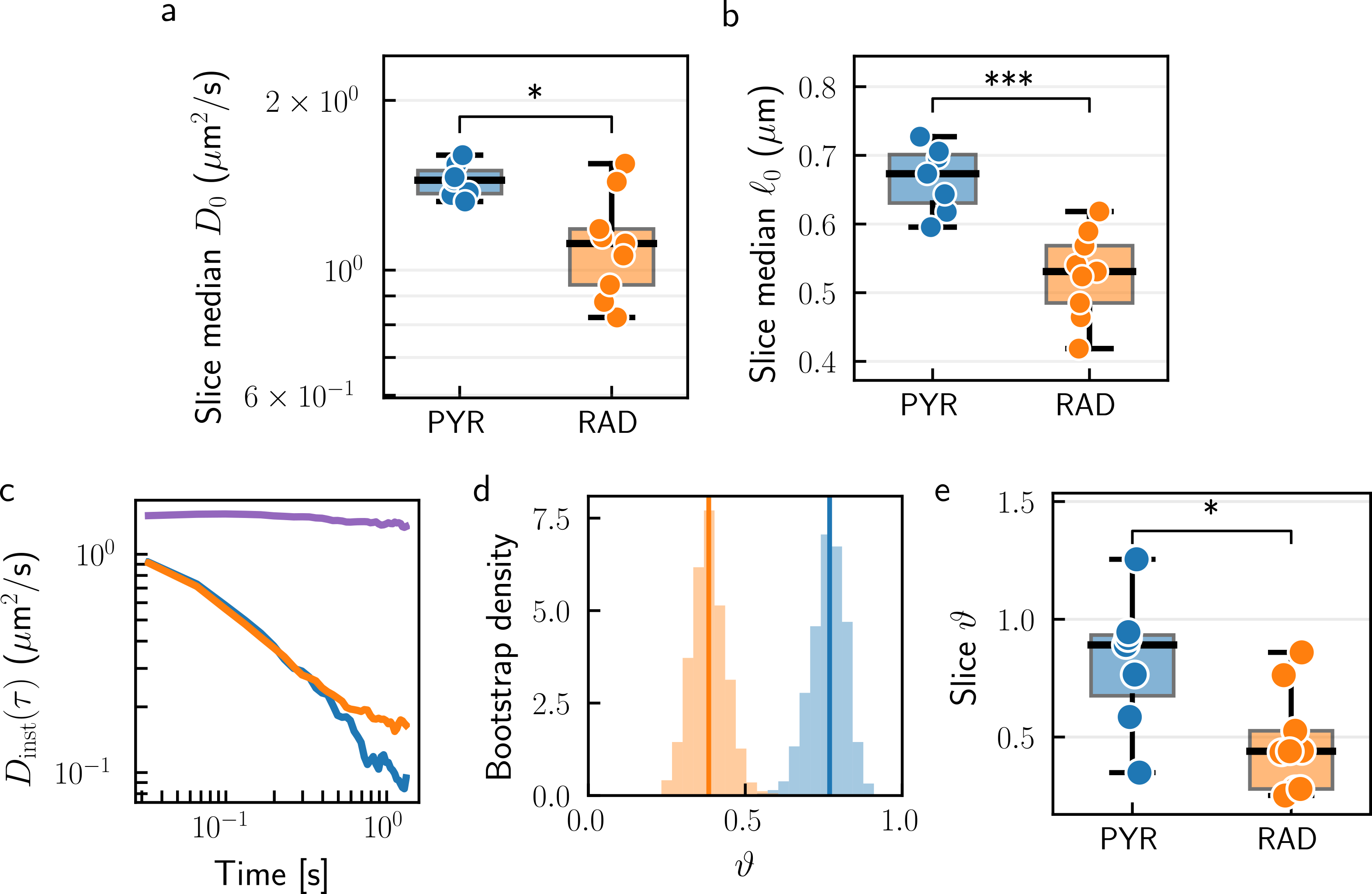}
\caption{\label{fig:regional_scale_dependent_transport}
\textbf{\emph{In-window} trajectories: Local mobility and subsequent transport hindrance differ between
hippocampal regions.}
All tissue analyses use \emph{in-window} trajectories.
(a) Slice-level median short-time diffusivity, $D_0$, estimated from the
initial Brownian-compatible regime preceding the crossover.
(b) Slice-level median crossover exploration length,
$\ell_0=\sqrt{\mathrm{MSD}(\tau_c)/3}$.
(c) Instantaneous diffusivity,
$D_{\mathrm{inst}}(\tau)=
\frac{1}{6}\mathrm{d}\langle\Delta r^2(\tau)\rangle/\mathrm{d}\tau$,
in PYR and RAD. Freely diffusing uCCNTs in water provide the experimental
reference.
(d) Slice-bootstrap distributions of the post-crossover restriction exponent,
$\vartheta$, obtained by fitting
$D_{\mathrm{inst}}(\tau)\propto\tau^{-\vartheta}$ after the crossover.
(e) Slice-level estimates of $\vartheta$.
In (a), (b), and (e), each point represents one slice containing more than
10 trajectories. Box plots show the median and interquartile range, with
whiskers extending to 1.5 times the interquartile range. Statistical
annotations report two-sided Mann--Whitney tests without multiple-comparison
correction; $^{*}P<0.05$ and $^{***}P<0.001$.
}
\end{figure*}

\noindent\textbf{ Local mobility and scale-dependent transport hindrance
distinguish hippocampal regions.}
Having established the dynamical changes associated with a resolved crossover,
we next asked whether their characteristic scales differed between
hippocampal regions. To account for biological variability, we pooled
trajectories within individual slices and included only slices containing
more than 10 \emph{in-window} trajectories (Supplementary Table~S2).

We first quantified mobility in the initial Brownian-compatible regime. The
slice-level median short-time diffusivity, $D_0$, was higher in PYR than in
RAD (Fig.~\ref{fig:regional_scale_dependent_transport}a; Mann--Whitney test,
$P<0.05$). We then defined the crossover exploration length from the measured
mean squared displacement at the crossover time,
$
\ell_0
=
\sqrt{\frac{\mathrm{MSD}(\tau_c)}{3}}.
$
The slice-level median $\ell_0$ was also larger in PYR than in RAD
(Fig.~\ref{fig:regional_scale_dependent_transport}b; Mann--Whitney test,
$P<0.001$). These trajectories therefore began with greater mobility and
explored a larger characteristic distance before transport hindrance became
detectable in the pyramidal layer.

To determine how transport evolved beyond the crossover, we calculated the
instantaneous diffusivity,
$
D_{\mathrm{inst}}(\tau)
=
\frac{1}{6}
\frac{\mathrm{d}\langle\Delta r^2(\tau)\rangle}
{\mathrm{d}\tau}.
$
The water reference remained approximately constant, whereas
$D_{\mathrm{inst}}(\tau)$ decreased strongly with lag time in both hippocampal
regions (Fig.~\ref{fig:regional_scale_dependent_transport}c). We quantified
this decline by fitting the post-crossover regime with
$
D_{\mathrm{inst}}(\tau)
\propto
\tau^{-\vartheta}.
$
Slice-bootstrap estimates gave a median $\vartheta$ of $0.77$ in PYR
(95\% confidence interval, $0.64$--$0.86$) and $0.38$ in RAD
($0.27$--$0.50$; Fig.~\ref{fig:regional_scale_dependent_transport}d).
Estimating $\vartheta$ independently for each eligible slice produced the
same regional ordering, with a more pronounced decline in diffusivity in PYR
than in RAD (Fig.~\ref{fig:regional_scale_dependent_transport}e;
Mann--Whitney test, $P<0.05$).

PYR thus combined greater initial mobility with a stronger subsequent decline
in diffusivity. A uniform multiplicative hindrance factor could change the
magnitude of diffusivity but would preserve its dependence on scale and
therefore could not produce the observed regional difference in $\vartheta$.
Our observations therefore indicate that extracellular transport differs
between these regions in both its initial mobility and its subsequent
evolution with scale, rather than through a single scale-independent rescaling
of diffusivity.\\

\begin{figure*}[htbp!]
\centering
\includegraphics[scale=1]{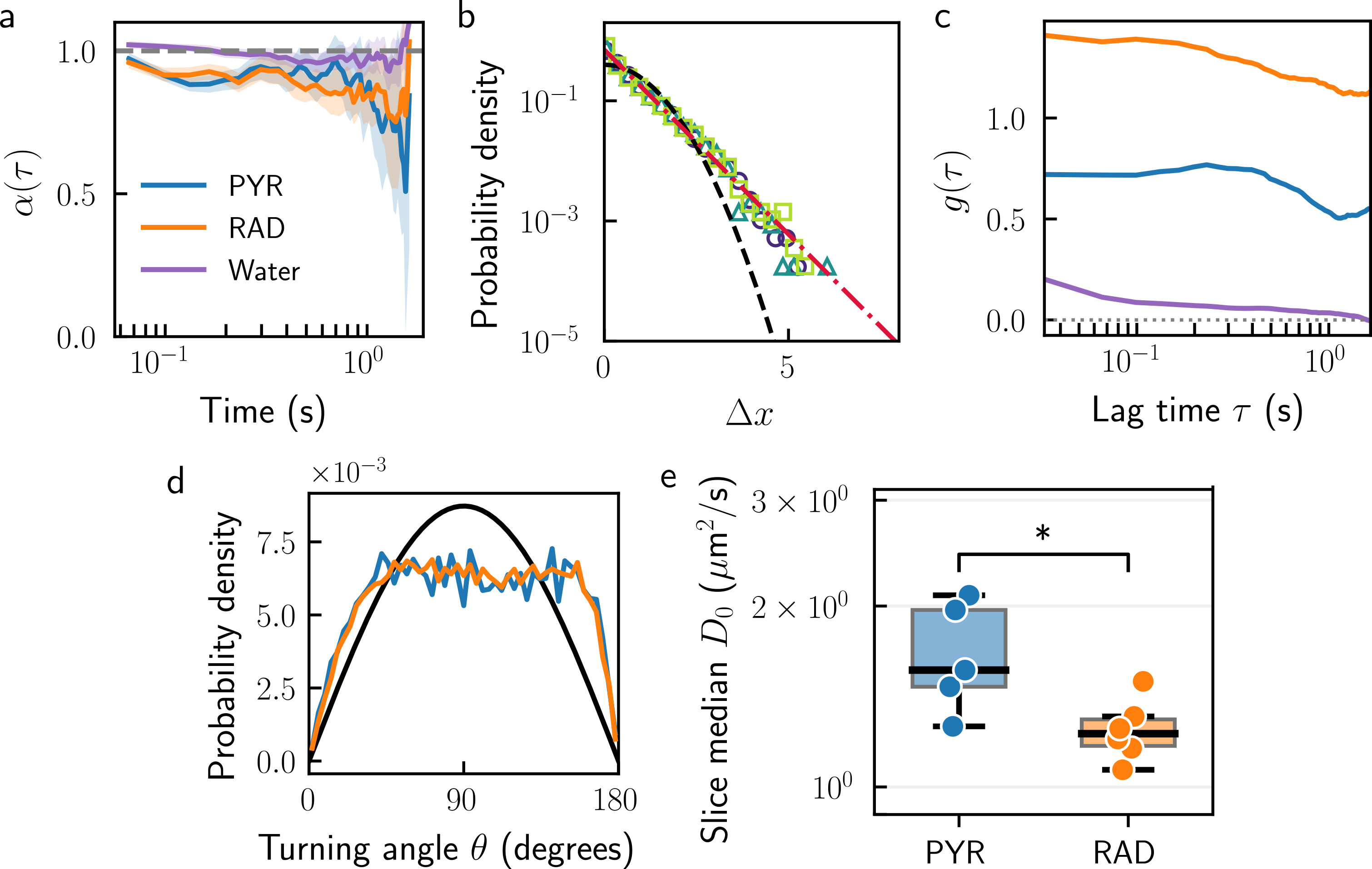}
\caption{\label{fig:brownian_retained_dynamics}
\textbf{Linear mean squared displacement scaling coexists with non-Gaussian
and directionally correlated motion.}
Analyses are shown for \emph{above-window} trajectories.
(a) Local scaling exponent,
$\alpha(\tau)=
\mathrm{d}\log\langle\Delta r^2(\tau)\rangle/
\mathrm{d}\log\tau$,
calculated from the time-ensemble-averaged mean squared displacement. The
dashed grey line marks the Brownian expectation, $\alpha=1$, and shaded
regions indicate 95\% trajectory-bootstrap confidence intervals.
(b) Ensemble-normalized self-van Hove distributions of one-dimensional
displacements for PYR trajectories at lag times
$\tau/\Delta t=1$, 3, and 5. Dashed black and dash-dotted red curves denote
unit-variance Gaussian and Laplace reference distributions, respectively.
(c) Lag-time dependence of the non-Gaussian parameter, $g(\tau)$, for PYR,
RAD, and water trajectories. The dotted grey line marks the Gaussian
expectation, $g=0$.
(d) Distributions of three-dimensional turning angles between consecutive
displacement vectors in PYR and RAD. The black curve denotes the isotropic
three-dimensional expectation,
$P(\theta)=\frac{1}{2}\sin\theta$.
(e) Slice-level median short-time diffusivity, $D_0$, of
\emph{above-window} trajectories in PYR and RAD. Box plots show the median
and interquartile range, with whiskers extending to 1.5 times the
interquartile range. The statistical annotation reports a two-sided
Mann--Whitney test performed at the slice level; $^{*}P<0.05$.
}
\end{figure*}

\pagebreak
\noindent\textbf{\emph{Above-window} trajectories retain Brownian MSD scaling
despite heterogeneous and directionally correlated dynamics.}
At one limit of the observed behaviour, some trajectories retained linear mean
squared displacement growth throughout the experimental window. This
population represented 19.4\% of PYR trajectories and 22.1\% of RAD
trajectories. The absence of a detected crossover does not establish that one
could not occur at larger scales. It instead identifies trajectories for which
a constant diffusivity adequately described second-moment growth over the
accessible interval. We therefore asked whether their motion also satisfied
the other predictions of isotropic Brownian diffusion: Gaussian, independent
increments without directional correlations
\cite{metzlerAnomalousDiffusionModels2014,
rehfeldtRandomWalkerToolbox2023}.
Their local scaling exponent remained close to unity over most of the measured
interval in PYR, RAD, and water
(Fig.~\ref{fig:brownian_retained_dynamics}a).

We next asked whether linear mean squared displacement growth was accompanied
by Gaussian displacement statistics. To compare distribution shapes across
lag times, we normalized displacements by their lag-dependent ensemble root
mean squared value. The resulting self-van Hove distributions collapsed onto
a common heavy-tailed profile, with a shape close to a Laplace distribution
(Fig.~\ref{fig:brownian_retained_dynamics}b)
\cite{wangAnomalousYetBrownian2009,
rehfeldtRandomWalkerToolbox2023}.
Accordingly, the non-Gaussian parameter, $g(\tau)$, remained positive in both
hippocampal regions while approaching zero in water
(Fig.~\ref{fig:brownian_retained_dynamics}c). Because population-level heavy
tails can arise from differences in diffusivity between otherwise Brownian
trajectories
\cite{wangAnomalousYetBrownian2009},
we repeated the analysis after normalizing each trajectory by its own root
mean squared displacement (Fig.~S3). Heavy tails persisted after this
normalization, showing that differences in constant diffusivity between
otherwise Brownian trajectories were insufficient to explain the observed
distributions.

The three-dimensional turning-angle distributions also differed markedly
from the isotropic expectation,
$
P(\theta)
=
\frac{1}{2}\sin\theta.
$
The reduced probability near $90^\circ$ and excess probability towards both
small and large angles indicate preferential continuation and reversal of
motion (Fig.~\ref{fig:brownian_retained_dynamics}d). These directional
correlations cannot arise from a static distribution of independent isotropic
Brownian trajectories, irrespective of the distribution of their diffusion
coefficients.

Finally, the slice-level median short-time diffusivity, $D_0$, was higher in
PYR than in RAD
(Fig.~\ref{fig:brownian_retained_dynamics}e; Mann--Whitney test,
$P<0.05$). Although $D_0$ was lower in the \emph{in-window} population than
in the corresponding \emph{above-window} population, the same regional
ordering, PYR above RAD, was retained.\\

In the next section, we finally examined the opposite observational limit, in
which departure from Brownian MSD scaling was already present at the earliest
accessible scale.\\

\noindent\textbf{\emph{Below-window} trajectories reveal strongly
anticorrelated dynamics at the earliest experimental scale.}
This early departure could arise from geometrical hindrance, confinement, or
obstruction within the extracellular space, but also from transient or
nonspecific interactions between uCCNTs and their surroundings
\cite{sykovaDiffusionBrainExtracellular2008a,
nicholsonBrainExtracellularSpace2017a}.
Because these mechanisms can produce similar mean squared displacement
signatures over a finite interval
\cite{hoflingAnomalousTransportComplex2013,
metzlerAnomalousDiffusionModels2014,
rehfeldtRandomWalkerToolbox2023},
the trajectories establish that motion was already altered at the earliest
accessible scale but do not identify the physical origin of that alteration.

\begin{figure}[htbp!]
\centering
\includegraphics[scale=1]{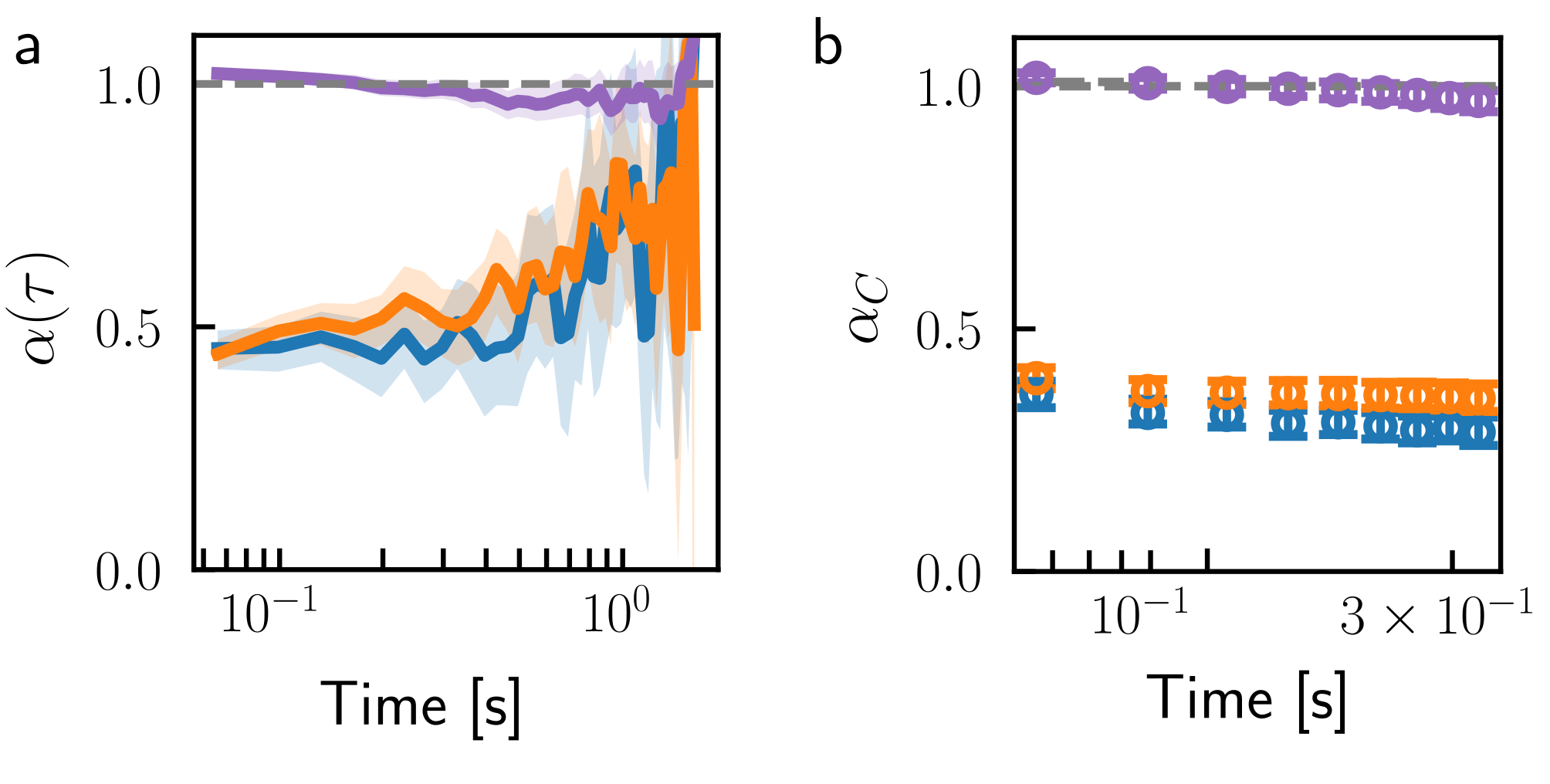}
\caption{\label{fig:immediate_departure_dynamics}
\textbf{Subdiffusive and antipersistent dynamics of \emph{below-window}
trajectories from the earliest resolved scales.}
(a) Local mean squared displacement scaling exponent for
\emph{below-window} trajectories in the pyramidal layer (PYR, blue) and
stratum radiatum (RAD, orange), with freely diffusing uCCNTs in water as the
experimental reference (purple).
(b) Increment-correlation exponent, $\alpha_C$, obtained by fitting the
normalized VACF over displacement intervals $\delta t$. Shaded regions and
error bars indicate 95\% trajectory-bootstrap confidence intervals.
}
\end{figure}

Consistent with this classification, the local scaling exponent,
$\alpha(\tau)$, was already below unity at the shortest resolved lag times in
PYR and RAD, in contrast to the water control
(Fig.~\ref{fig:immediate_departure_dynamics}a). The increment-correlation
exponent, $\alpha_C$, was likewise substantially below unity from the earliest
displacement interval and decreased further as that interval increased
(Fig.~\ref{fig:immediate_departure_dynamics}b). The \emph{below-window}
category was therefore characterized by sublinear mean squared displacement
growth and antipersistent motion from the beginning of the experimental
window
\cite{metzlerAnomalousDiffusionModels2014,
rehfeldtRandomWalkerToolbox2023}.

Because no preceding Brownian-compatible regime was resolved, neither the
pre-alteration diffusivity, $D_0$, nor the crossover time, $\tau_c$, could be
measured directly. A diffusivity estimated from the earliest available
displacements would already incorporate unresolved constraints or
interactions and should therefore be interpreted as an effective
post-alteration value, rather than the mobility preceding them.

The three operational categories can be placed within a common observational
framework, while recognizing that this framework does not establish a shared
mechanism. For \emph{in-window} trajectories, both the initial
Brownian-compatible regime and the subsequent departure were observed. Under
a unified crossover interpretation, \emph{below-window} trajectories would
represent left-censored crossover times occurring before the earliest
resolved scale, whereas \emph{above-window} trajectories would represent
right-censored crossover times lying beyond the measured interval, if a
crossover occurs at all. The categories may therefore reflect different
positions of a distributed crossover scale relative to the observation
window, rather than intrinsically distinct modes of motion.

\emph{Below-window} trajectories comprised 34.1\% of PYR trajectories and
37.0\% of RAD trajectories. Transport was therefore already modified at the
earliest accessible scale for approximately one third of extracellular
trajectories. Together with the directly resolved crossovers of
\emph{in-window} trajectories and the retained Brownian MSD scaling of
\emph{above-window} trajectories, this result shows why a single diffusion
coefficient measured over one interval cannot generally describe
extracellular exploration across scales.

\section*{Discussion}

Our trajectory-resolved measurements show that extracellular transport in the
living brain evolves continuously as molecules explore larger distances.
Individual probes may move rapidly within their immediate surroundings before
encountering restrictions imposed by structural boundaries, narrow
connections, or the composition of the ECS itself. Because these features vary
throughout the extracellular space, the scale at which they alter transport
also varies between trajectories
\cite{sykovaDiffusionBrainExtracellular2008a,
kinneyExtracellularSheetsTunnels2013,
godinSingleNanotubeTrackingReveals2017a,
grassiNanoscaleFunctionalHeterogeneity2023}.
This variation suggests a unified interpretation of the three trajectory
categories: the MSD crossover occurs below the measured window for
below-window trajectories, within it for in-window trajectories, and beyond it
for above-window trajectories. Geometrical confinement, restricted
connectivity, and transient molecular interactions can each contribute to this
crossover behavior. The three categories may therefore reflect different
positions of a common phenomenon relative to the observation window rather
than intrinsically distinct transport modes.

The presented framework provides three biologically informative observables
extracted directly from single trajectories: a short-time diffusivity $D_0$, a
crossover length $\ell_0$ marking the scale at which transport hindrance becomes
detectable, and a post-crossover diffusivity-decay exponent $\vartheta$
characterizing its subsequent evolution. Median crossover lengths of
$0.5$--$0.7~\mu\mathrm{m}$ place the detected transition within the
submicrometric range of extracellular exploration, comparable to reported
submicrometric structural length scales in the ECS. These observables reveal a
decoupling of local mobility from large-scale exploration between hippocampal
layers: PYR combines a higher short-time diffusivity with a stronger subsequent
decline in instantaneous diffusivity than RAD. This pattern is consistent with
PYR containing locally permissive spaces separated by boundaries or connections
that limit larger-scale exploration, whereas RAD may hinder motion more
continuously. The two regional signatures are incompatible with a uniform,
single-parameter hindrance model, irrespective of the underlying
microstructural mechanisms.

Although future work will be needed to identify the underlying mechanisms, for
example, by varying probe size or surface chemistry
\cite{nanceDensePolyEthyleneGlycol2012,
etocNonspecificInteractionsGovern2018},
or by correlating tracer trajectories with ECS ultrastructure, our findings
provide a framework for understanding a longstanding tension in the field.
Diffusion MRI, real-time iontophoresis, and single-particle tracking probe
different molecular species across distinct spatial and temporal regimes
\cite{nicholsonExtracellularSpaceStructure1998,
sykovaDiffusionBrainExtracellular2008a,
novikovQuantifyingBrainMicrostructure2019,
godinSingleNanotubeTrackingReveals2017a}
and have historically yielded inconsistent estimates of extracellular
transport parameters. Our results identify a general mechanism by which such
differences can arise: a diffusion coefficient that captures spreading within
one observational window does not need to predict transport beyond it. Probe size and
tissue organization determine which pathways through the extracellular network
are accessible, while interactions along those pathways further modulate motion
\cite{godinSingleNanotubeTrackingReveals2017a,
grassiNanoscaleFunctionalHeterogeneity2023,
thorneVivoDiffusionAnalysis2006}.
Effective diffusion coefficients are therefore not intrinsic constants of the
extracellular space, but scale- and probe-dependent observables of a multiscale
transport process. From this perspective, apparent methodological discrepancies
may instead represent complementary observations of the same underlying
phenomenon.

\appendix

\section*{Methods}

\begingroup
\fontsize{8.5}{10.2}\selectfont

\noindent\textbf{Organotypic hippocampal slice cultures.}
Cortico-hippocampal organotypic slice cultures were prepared from three
postnatal day 5--6 Sprague Dawley rats using a membrane-interface method
adapted from Stoppini et al.\
\cite{stoppiniSimpleMethodOrganotypic1991}.

After extraction, the brain hemispheres were separated and the
cortico-hippocampal complexes were isolated. The meninges were carefully
removed with fine forceps. Dissection was completed within approximately
20~min while the tissue was maintained in cold, sterile dissection medium
containing, in mM, 0.5 CaCl$_2$, 2.5 KCl, 2 MgCl$_2$,
0.66 KH$_2$PO$_4$, 0.85 Na$_2$HPO$_4$, 0.28 MgSO$_4$, 50 NaCl,
2.7 NaHCO$_3$, 25 glucose, 175 sucrose, 2 HEPES, and 0.1 kynurenic acid
to limit excitotoxicity (all reagents from Sigma-Aldrich unless otherwise
specified).

Transverse slices, 350~$\mu$m thick, were prepared using a tissue chopper
(model TC752, McIlwain). Slices with preserved hippocampal morphology and
tissue integrity were selected and allowed to recover for 20~min at
4~$^\circ$C in dissection medium without kynurenic acid. Individual slices
were transferred onto sterilized $6\times6$~mm pieces of hydrophilic
polytetrafluoroethylene membrane (FHLC04700, Merck Millipore), positioned on
Millicell culture inserts with 0.4~$\mu$m pores and a diameter of 30~mm
(Merck Millipore). This support facilitated the handling of individual slices
while limiting deformation of extracellular-space architecture
\cite{grassiNanoscaleFunctionalHeterogeneity2023}.

The inserts were placed in six-well plates containing 1~mL of preheated
culture medium per well. The culture medium consisted of 50\% Basal Medium
Eagle, 25\% Hank's Balanced Salt Solution, 25\% horse serum, 11.2~mM glucose,
20~mM GlutaMAX, and 0.8~mM NaHCO$_3$ (all products from Gibco unless
otherwise specified). Four slices were maintained on each insert at
35~$^\circ$C, 5\% CO$_2$, and greater than 90\% relative humidity. The
culture medium was completely replaced on day in vitro 1 and every two days
thereafter. Experiments were performed after 7--10 days in vitro.

\noindent\textbf{Ultrashort carbon nanotubes.}
uCCNTs were prepared following the protocol reported previously
\cite{nandiUltrashortCarbonNanotubes2025b}. Briefly, monochiral
$(6,5)$ single-walled carbon nanotubes were sorted by aqueous two-phase
extraction as described previously
\cite{liSeparationSmallDiameterSingleWalled2019c}
and shortened by extended tip sonication. They were then functionalized with
luminescent oxygen defects through a Fenton-like reaction using copper(II)
sulfate pentahydrate and sodium L-ascorbate according to an established
protocol
\cite{setteleEasyAccessBright2024}.
This procedure was followed by surfactant exchange to PLPEG5000
(Avanti Polar Lipids) by dialysis.

\noindent\textbf{Near-infrared three-dimensional single-particle tracking.}
Near-infrared photoluminescence imaging of
phospholipid--polyethylene-glycol-functionalized uCCNTs was performed using a
custom-built inverted microscope based on a Nikon Eclipse Ti platform. The
microscope was equipped with a $60\times$ water-immersion objective with a
numerical aperture of 1.27 (Nikon). The uCCNTs were excited using a 988~nm
diode laser (AeroDIODE), resonant with the first excitonic transition of
$(6,5)$ nanotubes. The excitation beam was circularly polarized to minimize
variations in excitation efficiency associated with the in-plane orientation
of individual nanotubes.

Photoluminescence emission was collected through the same objective and
detected using an InGaAs camera (C-RED 2, First Light Imaging). A dichroic
mirror (Di02-R1064, Semrock) directed the excitation beam towards the sample,
and a 1064~nm long-pass filter (RazorEdge 1064, Semrock) suppressed reflected
excitation light while transmitting the red-shifted uCCNT emission.

Three-dimensional localization was achieved using a double-helix
point-spread function (DH-PSF). A relay system composed of two achromatic
doublet lenses in a $4f$ configuration was incorporated into the detection
path. A double-helix phase mask (Double Helix Optics) was positioned in the
Fourier plane of the relay system, optically conjugate to the back focal plane
of the objective. The phase mask was mounted on a kinematic stage with an
adjustment precision of approximately 1~$\mu$m to permit accurate alignment
of the phase modulation.

\noindent\textbf{DH-PSF calibration and localization precision.}
The experimentally generated DH-PSF was calibrated using immobilized uCCNTs.
Glass coverslips were plasma-cleaned at 180~W for 2~min and incubated with
100~$\mu$L of a 0.1~g/L aqueous poly-L-lysine solution for 30~min. A
50~$\mu$L aliquot of diluted uCCNT dispersion was then deposited onto the
coated coverslip and incubated for 1~h. Unbound nanotubes were removed by
rinsing with deionized water, and the samples were rewetted with deionized
water immediately before calibration.

An isolated, bright, and photostable uCCNT was selected for calibration.
Axial image stacks were acquired in 50~nm steps over a range of approximately
5~$\mu$m. At each axial position, 50 images were recorded with an exposure
time of 33~ms and averaged to improve the signal-to-noise ratio. The
resulting experimental axial stack was used for phase retrieval in ZOLA
\cite{aristovZOLA3DAllowsFlexible2018b},
using 50 optimization iterations. Three-dimensional localization was
subsequently performed in ZOLA using the experimentally retrieved PSF model.

Localization precision was evaluated directly from the 50 individual images
acquired at each axial position. For each coordinate, precision was defined
as the standard deviation of the repeated position estimates. The detected
signal and background for the representative calibration were approximately
35,000 and 3 photoelectrons, respectively. The corresponding localization
precision was approximately 5~nm in the lateral dimensions and 10~nm in the
axial dimension near the centre of the calibrated range, with
position-dependent precision and Cramér--Rao lower bounds reported in
Fig.~S1.

\noindent\textbf{uCCNT loading and trajectory acquisition in organotypic
tissue.}
After 7--10 days in vitro, slices were incubated with uCCNTs for 3~h in
culture medium at 35~$^\circ$C, 5\% CO$_2$, and greater than 90\% relative
humidity. Before imaging, loaded slices were washed for 10~min in saline
containing, in mM, 145 NaCl, 4 KCl, 2 CaCl$_2$, 0.5 MgCl$_2$, 10 HEPES,
and 10 glucose, adjusted to pH~7.4 and approximately 300~mOsm.

Individual organotypic slices were transferred to the imaging chamber and
maintained in the same saline solution. PYR and RAD recording regions were
identified from their anatomical appearance under transmitted white-light
imaging. Near-infrared image sequences were acquired with an exposure time of
33~ms for 3000 frames. The excitation power density at the sample was
approximately $3~\mathrm{kW\,cm^{-2}}$. To minimize contributions from
damaged or structurally altered tissue near the exposed surface, all uCCNT
trajectories were recorded at least 10~$\mu$m below the slice surface.
Recordings were obtained from seven PYR slices and nine RAD slices derived
from three animals.

Individual uCCNT positions were localized in three dimensions using ZOLA and
the experimentally retrieved DH-PSF. Localizations were linked into
trajectories using the TrackPy Python package with a maximum frame-to-frame
displacement of 3 pixels, a maximum gap of 2 frames, and a minimum accepted
trajectory length of 50 frames.

\noindent\textbf{Mean squared displacement and Brownian compatibility.}
Three-dimensional trajectories were represented by Cartesian coordinates
$\mathbf{r}(t)=[x(t),y(t),z(t)]$ sampled at $\Delta t=0.033$~s. For each
trajectory, the time-averaged mean squared displacement was calculated as
\[
\overline{\delta^2(\tau)}
=
\left\langle
\left|
\mathbf{r}(t+\tau)-\mathbf{r}(t)
\right|^2
\right\rangle_t
\]
for lag numbers 1 through 50, corresponding to 0.033--1.650~s.

Compatibility with Brownian mean squared displacement scaling was evaluated
by fitting
\[
\overline{\delta^2(\tau)}=6D\tau
\]
over progressively increasing lag-time windows. At each window length, the
relative fitting error was compared with the corresponding empirical
distribution obtained from freely diffusing uCCNTs in water. The 95th
percentile of the water distribution defined the acceptance threshold. Water
trajectories were evaluated by leave-one-trajectory-out calibration. The
window-dependent error distributions and thresholds are reported in
Fig.~S2.

Departure was considered irreversible after a trajectory first exceeded the
water-calibrated threshold. Trajectories were classified as
\emph{below-window} when incompatibility occurred at the earliest tested
window, \emph{in-window} when an initially compatible trajectory subsequently
departed within the observation window, or \emph{above-window} when
compatibility persisted through all 50 lag points. These categories describe
the position of the detected departure relative to the experimentally
accessible interval and do not imply intrinsically distinct particle types.
Full definitions and validation of the classification are provided in the
Supplementary Methods.

\noindent\textbf{Local mobility and crossover parameters.}
The short-time diffusivity, $D_0$, was estimated from a zero-intercept fit to
the first three mean squared displacement points for trajectories with a
resolved initial Brownian-compatible regime. For \emph{in-window}
trajectories, the crossover time, $\tau_c$, was defined by the first
irreversible failure of the Brownian compatibility test. The corresponding
crossover exploration length was calculated from the measured mean squared
displacement as
\[
\ell_0
=
\sqrt{
\frac{
\overline{\delta^2(\tau_c)}
}{3}
}.
\]
Values of $\tau_c$ and $\ell_0$ were not assigned to \emph{below-window} or
\emph{above-window} trajectories. Because \emph{below-window} trajectories
lacked a resolved Brownian-compatible regime, their earliest apparent
diffusivity was interpreted as resolution limited rather than as an estimate
of the unconstrained microscopic diffusivity.

\noindent\textbf{Scale-dependent transport.}
For each condition and operational category, trajectory mean squared
displacement curves were averaged at each lag time. The local scaling
exponent was calculated as
\[
\alpha(\tau)
=
\frac{
\mathrm{d}\log
\left\langle\overline{\delta^2(\tau)}\right\rangle
}{
\mathrm{d}\log\tau
},
\]
where $\alpha=1$ corresponds to linear mean squared displacement growth.

Instantaneous diffusivity was calculated as
\[
D_{\mathrm{inst}}(\tau)
=
\frac{1}{6}
\frac{
\mathrm{d}
\left\langle\overline{\delta^2(\tau)}\right\rangle
}{
\mathrm{d}\tau
}.
\]
For \emph{in-window} trajectories, post-crossover restriction was quantified
by fitting
\[
D_{\mathrm{inst}}(\tau)=A\tau^{-\vartheta}
\]
over lag numbers 7 through 40. This fitting interval was selected without
reference to condition from the sensitivity analysis reported in Fig.~S4.

\noindent\textbf{Displacement and directional statistics.}
Displacement statistics were characterized using normalized self-van Hove
distributions and the non-Gaussian parameter
\[
g(\tau)
=
\frac{
\left\langle\Delta r_k^4(\tau)\right\rangle
}{
3\left\langle\Delta r_k^2(\tau)\right\rangle^2
}
-1.
\]
Displacements were normalized either by their ensemble root mean squared value
or, as a control, by the lag-dependent root mean squared displacement of each
trajectory before pooling. The trajectory-normalized control is reported in
Fig.~S3. Directional correlations were evaluated from three-dimensional
turning angles between successive displacement vectors and from normalized
increment-autocorrelation functions. Detailed definitions, null
distributions, and fitting procedures are provided in the Supplementary
Methods.

\noindent\textbf{Uncertainty estimation and statistical analysis.}
Uncertainty in pooled trajectory statistics was estimated by resampling
complete trajectories with replacement. Uncertainty in the post-crossover
restriction exponent was estimated using a slice bootstrap in which eligible
slices were sampled with replacement and the complete calculation of the
ensemble mean squared displacement, instantaneous diffusivity, and
$\vartheta$ was repeated.

Regional comparisons were performed at the slice level. Trajectory-derived
parameters were summarized by their median within each slice, and
slice--category groups were included only when they contained strictly more
than 10 valid trajectories. PYR and RAD were compared using two-sided
Mann--Whitney $U$ tests, with statistical significance defined as $P<0.05$.
Slices were obtained from three animals. Analyses were performed in Python
using NumPy, pandas, SciPy, and TrackPy.

\endgroup



\begin{acknowledgments}

We thank M. Tondusson for experimental assistance and support. We thank the
Cell Biology Facility for supporting cell-biology-related activities,
particularly Morgane Meras and Delphine Bouchet. L.C. and L.G. acknowledge
financial support from the European Research Council Synergy Grant 951294.
L.C. acknowledges support from the Agence Nationale de la Recherche
(EUR Light\&T, PIA3 Program, ANR-17-EURE-0027) and IdEx Bordeaux
(Grand Research Program GPR LIGHT). F.L.S. and J.Z. acknowledge funding from
the European Research Council under the European Union's Horizon 2020 research
and innovation programme (Grant Agreement No. 817494, ``TRIFECTS''). J.Z.
acknowledges additional support from the Deutsche Forschungsgemeinschaft
(DFG, German Research Foundation) under Germany's Excellence Strategy for the
Cluster of Excellence ``3D Matter Made to Order''
(EXC-2082/1, project 390761711).

\end{acknowledgments}

\bibliography{bib}

@article{metzlerAnomalousDiffusionModels2014,
  title = {Anomalous Diffusion Models and Their Properties: Non-Stationarity, Non-Ergodicity, and Ageing at the Centenary of Single Particle Tracking},
  shorttitle = {Anomalous Diffusion Models and Their Properties},
  author = {Metzler, Ralf and Jeon, Jae-Hyung and Cherstvy, Andrey G. and Barkai, Eli},
  year = 2014,
  journal = {Phys. Chem. Chem. Phys.},
  volume = {16},
  number = {44},
  pages = {24128--24164},
  issn = {1463-9076, 1463-9084},
  doi = {10.1039/C4CP03465A},
  urldate = {2024-01-09},
  langid = {english}
}

@article{thorneInVivoDiffusionAnalysis2006,
  author  = {Thorne, Robert G. and Nicholson, Charles},
  title   = {In vivo diffusion analysis with quantum dots and dextrans predicts the width of brain extracellular space},
  journal = {Proceedings of the National Academy of Sciences},
  year    = {2006},
  volume  = {103},
  number  = {14},
  pages   = {5567--5572},
  doi     = {10.1073/pnas.0509425103}
}

@article{hoflingAnomalousTransportComplex2013,
  author  = {H{\"o}fling, Felix and Franosch, Thomas},
  title   = {Anomalous transport in the crowded world of biological cells},
  journal = {Reports on Progress in Physics},
  year    = {2013},
  volume  = {76},
  number  = {4},
  pages   = {046602},
  doi     = {10.1088/0034-4885/76/4/046602}
}

@article{rehfeldtRandomWalkerToolbox2023,
  author  = {Rehfeldt, Florian and Weiss, Matthias},
  title   = {The random walker's toolbox for analyzing single-particle tracking data},
  journal = {Soft Matter},
  year    = {2023},
  volume  = {19},
  number  = {27},
  pages   = {5206--5222},
  doi     = {10.1039/D3SM00557G}
}

@article{wangAnomalousYetBrownian2009,
  author  = {Wang, Bin and Anthony, Stephen M. and Bae, Sung Chul and Granick, Steve},
  title   = {Anomalous yet Brownian},
  journal = {Proceedings of the National Academy of Sciences},
  year    = {2009},
  volume  = {106},
  number  = {36},
  pages   = {15160--15164},
  doi     = {10.1073/pnas.0903554106}
}

@article{aristovZOLA3DAllowsFlexible2018b,
  title = {{{ZOLA-3D}} Allows Flexible {{3D}} Localization Microscopy over an Adjustable Axial Range},
  author = {Aristov, Andrey and Lelandais, Benoit and Rensen, Elena and Zimmer, Christophe},
  year = 2018,
  month = jun,
  journal = {Nature Communications},
  volume = {9},
  number = {1},
  pages = {2409},
  publisher = {Nature Publishing Group},
  issn = {2041-1723},
  doi = {10.1038/s41467-018-04709-4},
  urldate = {2025-03-04},
  copyright = {2018 The Author(s)},
  langid = {english}
}

@article{etocNonspecificInteractionsGovern2018,
  title = {Non-Specific Interactions Govern Cytosolic Diffusion of Nanosized Objects in Mammalian Cells},
  author = {Etoc, Fred and Balloul, Elie and Vicario, Chiara and Normanno, Davide and Li{\ss}e, Domenik and Sittner, Assa and Piehler, Jacob and Dahan, Maxime and Coppey, Mathieu},
  year = 2018,
  month = aug,
  journal = {Nature Materials},
  volume = {17},
  number = {8},
  pages = {740--746},
  issn = {1476-1122, 1476-4660},
  doi = {10.1038/s41563-018-0120-7},
  urldate = {2024-11-13},
  langid = {english}
}

@article{gahwilerOrganotypicCulturesNeural1988,
  title = {Organotypic Cultures of Neural Tissue},
  author = {G{\"a}hwiler, B. H.},
  year = 1988,
  month = jan,
  journal = {Trends in Neurosciences},
  volume = {11},
  number = {11},
  pages = {484--489},
  publisher = {Elsevier},
  issn = {0166-2236, 1878-108X},
  doi = {10.1016/0166-2236(88)90007-0},
  urldate = {2026-03-17},
  langid = {english},
  pmid = {2469173}
}

@article{godinSingleNanotubeTrackingReveals2017a,
  title = {Single-{{Nanotube Tracking Reveals}} the {{Nanoscale Organization}} of the {{Extracellular Space}} in the {{Live Brain}}},
  author = {Godin, A. G. and Varela, J. A. and Gao, Z. and Dann{\'e}, N. and Dupuis, J. P. and Lounis, B. and Groc, L. and Cognet, L.},
  year = 2017,
  journal = {Nat. Nanotechnol.},
  volume = {12},
  number = {3},
  pages = {238}
}

@article{grassiNanoscaleFunctionalHeterogeneity2023,
  title = {Nanoscale and Functional Heterogeneity of the Hippocampal Extracellular Space},
  author = {Grassi, Diego and Idziak, Agata and Lee, Antony and Calaresu, Ivo and Sibarita, Jean-Baptiste and Cognet, Laurent and N{\"a}gerl, U. Valentin and Groc, Laurent},
  year = 2023,
  month = may,
  journal = {Cell Reports},
  volume = {42},
  number = {5},
  publisher = {Elsevier},
  issn = {2211-1247},
  doi = {10.1016/j.celrep.2023.112478},
  urldate = {2024-10-24},
  langid = {english},
  pmid = {37149864}
}

@article{kinneyExtracellularSheetsTunnels2013,
  title = {Extracellular Sheets and Tunnels Modulate Glutamate Diffusion in Hippocampal Neuropil},
  author = {Kinney, Justin P. and Spacek, Josef and Bartol, Thomas M. and Bajaj, Chandrajit L. and Harris, Kristen M. and Sejnowski, Terrence J.},
  year = 2013,
  journal = {Journal of Comparative Neurology},
  volume = {521},
  number = {2},
  pages = {448--464},
  issn = {1096-9861},
  doi = {10.1002/cne.23181},
  urldate = {2023-10-04},
  langid = {english}
}

@article{korogodUltrastructuralAnalysisAdult2015,
  title = {Ultrastructural Analysis of Adult Mouse Neocortex Comparing Aldehyde Perfusion with Cryo Fixation},
  author = {Korogod, Natalya and Petersen, Carl Ch and Knott, Graham W},
  year = 2015,
  month = aug,
  journal = {eLife},
  volume = {4},
  pages = {e05793},
  issn = {2050-084X},
  doi = {10.7554/eLife.05793},
  urldate = {2025-06-30},
  copyright = {http://creativecommons.org/licenses/by/4.0/},
  langid = {english}
}

@article{liSeparationSmallDiameterSingleWalled2019c,
  title = {Separation of {{Small-Diameter Single-Walled Carbon Nanotubes}} in {{One}} to {{Three Steps}} with {{Aqueous Two-Phase Extraction}}},
  author = {Li, Han and Gordeev, Georgy and Garrity, Oisin and Reich, Stephanie and Flavel, Benjamin S.},
  year = 2019,
  month = jan,
  journal = {ACS Nano},
  pages = {acsnano.8b09579},
  issn = {1936-0851, 1936-086X},
  doi = {10.1021/acsnano.8b09579},
  urldate = {2026-03-17},
  copyright = {https://doi.org/10.15223/policy-029},
  langid = {english}
}

@article{nanceDensePolyEthyleneGlycol2012,
  title = {A {{Dense Poly}}({{Ethylene Glycol}}) {{Coating Improves Penetration}} of {{Large Polymeric Nanoparticles Within Brain Tissue}}},
  author = {Nance, Elizabeth A. and Woodworth, Graeme F. and Sailor, Kurt A. and Shih, Ting-Yu and Xu, Qingguo and Swaminathan, Ganesh and Xiang, Dennis and Eberhart, Charles and Hanes, Justin},
  year = 2012,
  month = aug,
  journal = {Science Translational Medicine},
  volume = {4},
  number = {149},
  issn = {1946-6234, 1946-6242},
  doi = {10.1126/scitranslmed.3003594},
  urldate = {2025-08-12},
  langid = {english}
}

@article{nandiUltrashortCarbonNanotubes2025b,
  title = {Ultrashort {{Carbon Nanotubes}} with {{Luminescent Color Centers Are Bright NIR-II Nanoemitters}}},
  author = {Nandi, Somen and Gresil, Quentin and Lambert, Benjamin P. and Sebastian, Finn L. and Settele, Simon and Calaresu, Ivo and {Estaun-Panzano}, Juan and Lovisotto, Anna and Mazzocco, Claire and Flavel, Benjamin S. and Bezard, Erwan and Groc, Laurent and Zaumseil, Jana and Cognet, Laurent},
  year = 2025,
  month = may,
  journal = {ACS Nano},
  pages = {acsnano.5c02171},
  issn = {1936-0851, 1936-086X},
  doi = {10.1021/acsnano.5c02171},
  urldate = {2025-05-20},
  copyright = {https://doi.org/10.15223/policy-029},
  langid = {english}
}

@article{nicholsonBrainExtracellularSpace2017a,
  title = {Brain {{Extracellular Space}}: {{The Final Frontier}} of {{Neuroscience}}},
  shorttitle = {Brain {{Extracellular Space}}},
  author = {Nicholson, Charles and Hrab{\v e}tov{\'a}, Sabina},
  year = 2017,
  month = nov,
  journal = {Biophysical Journal},
  volume = {113},
  number = {10},
  pages = {2133--2142},
  issn = {00063495},
  doi = {10.1016/j.bpj.2017.06.052},
  urldate = {2024-01-27},
  langid = {english}
}

@article{nicholsonExtracellularSpaceStructure1998,
  title = {Extracellular Space Structure Revealed by Diffusion Analysis},
  author = {Nicholson, Charles and Sykov{\'a}, Eva},
  year = 1998,
  month = may,
  journal = {Trends in Neurosciences},
  volume = {21},
  number = {5},
  pages = {207--215},
  issn = {01662236},
  doi = {10.1016/S0166-2236(98)01261-2},
  urldate = {2025-08-04},
  copyright = {https://www.elsevier.com/tdm/userlicense/1.0/},
  langid = {english}
}

@article{novikovQuantifyingBrainMicrostructure2019,
  title = {Quantifying Brain Microstructure with Diffusion {{MRI}}: {{Theory}} and Parameter Estimation},
  shorttitle = {Quantifying Brain Microstructure with Diffusion {{MRI}}},
  author = {Novikov, Dmitry S. and Fieremans, Els and Jespersen, Sune N. and Kiselev, Valerij G.},
  year = 2019,
  month = apr,
  journal = {NMR in biomedicine},
  volume = {32},
  number = {4},
  pages = {e3998},
  issn = {0952-3480},
  doi = {10.1002/nbm.3998},
  urldate = {2026-01-15},
  pmcid = {PMC6481929},
  pmid = {30321478}
}

@article{pavaniThreedimensionalSinglemoleculeFluorescence2009,
  title = {Three-Dimensional, Single-Molecule Fluorescence Imaging beyond the Diffraction Limit by Using a Double-Helix Point Spread Function},
  author = {Pavani, Sri Rama Prasanna and Thompson, Michael A. and Biteen, Julie S. and Lord, Samuel J. and Liu, Na and Twieg, Robert J. and Piestun, Rafael and Moerner, W. E.},
  year = 2009,
  month = mar,
  journal = {Proceedings of the National Academy of Sciences},
  volume = {106},
  number = {9},
  pages = {2995--2999},
  issn = {0027-8424, 1091-6490},
  doi = {10.1073/pnas.0900245106},
  urldate = {2022-04-15},
  langid = {english}
}

@article{rehfeldtRandomWalkersToolbox2023,
  title = {The Random Walker's Toolbox for Analyzing Single-Particle Tracking Data},
  author = {Rehfeldt, Florian and Weiss, Matthias},
  year = 2023,
  journal = {Soft Matter},
  volume = {19},
  number = {28},
  pages = {5206--5222},
  issn = {1744-683X, 1744-6848},
  doi = {10.1039/D3SM00557G},
  urldate = {2024-01-09},
  langid = {english}
}

@article{setteleEasyAccessBright2024,
  title = {Easy {{Access}} to {{Bright Oxygen Defects}} in {{Biocompatible Single-Walled Carbon Nanotubes}} via a {{Fenton-like Reaction}}},
  author = {Settele, Simon and Stammer, Florian and Sebastian, Finn L. and Lindenthal, Sebastian and Wald, Simon R. and Li, Han and Flavel, Benjamin S. and Zaumseil, Jana},
  year = 2024,
  month = aug,
  journal = {ACS Nano},
  volume = {18},
  number = {31},
  pages = {20667--20678},
  issn = {1936-0851, 1936-086X},
  doi = {10.1021/acsnano.4c06448},
  urldate = {2025-03-24},
  copyright = {https://creativecommons.org/licenses/by-nc-nd/4.0/},
  langid = {english}
}

@article{soriaSynucleinopathyAltersNanoscale2020,
  title = {Synucleinopathy Alters Nanoscale Organization and Diffusion in the Brain Extracellular Space through Hyaluronan Remodeling},
  author = {Soria, Federico N. and Paviolo, Chiara and Doudnikoff, Evelyne and Arotcarena, Marie-Laure and Lee, Antony and Dann{\'e}, No{\'e}mie and Mandal, Amit Kumar and Gosset, Philippe and Dehay, Benjamin and Groc, Laurent and Cognet, Laurent and Bezard, Erwan},
  year = 2020,
  month = jul,
  journal = {Nature Communications},
  volume = {11},
  number = {1},
  pages = {3440},
  issn = {2041-1723},
  doi = {10.1038/s41467-020-17328-9}
}

@article{stoppiniSimpleMethodOrganotypic1991,
  title = {A Simple Method for Organotypic Cultures of Nervous Tissue},
  author = {Stoppini, L. and Buchs, P. A. and Muller, D.},
  year = 1991,
  month = apr,
  journal = {Journal of Neuroscience Methods},
  volume = {37},
  number = {2},
  pages = {173--182},
  issn = {0165-0270},
  doi = {10.1016/0165-0270(91)90128-m},
  langid = {english},
  pmid = {1715499}
}

@article{sykovaDiffusionBrainExtracellular2008a,
  title = {Diffusion in {{Brain Extracellular Space}}},
  author = {Sykov{\'a}, Eva and Nicholson, Charles},
  year = 2008,
  month = oct,
  journal = {Physiological Reviews},
  volume = {88},
  number = {4},
  pages = {1277--1340},
  issn = {0031-9333, 1522-1210},
  doi = {10.1152/physrev.00027.2007},
  urldate = {2025-07-31},
  langid = {english}
}

@article{sykovaDisruptionExtracellularMatrix2025,
  title = {Disruption of {{Extracellular Matrix}} and {{Perineuronal Nets Modulates Extracellular Space Volume}} and {{Geometry}}},
  author = {Sykov{\'a}, Eva and Vo{\v r}{\'i}{\v s}ek, Ivan and Star{\v c}uk, Zenon and Kratochv{\'i}la, Ji{\v r}{\'i} and Pavlova, Iveta and Ichikawa, Yuki and Kwok, Jessica C. F. and Kmon{\'i}{\v c}kov{\'a}, Eva and Myronchenko, Svitlana and Hrom{\'a}dka, Tom{\'a}{\v s} and Smolek, Tom{\'a}{\v s} and Avila, Martin and Basheer, Neha and {\v Z}ilka, Norbert},
  year = 2025,
  month = feb,
  journal = {The Journal of Neuroscience},
  volume = {45},
  number = {8},
  pages = {e0517242024},
  issn = {0270-6474, 1529-2401},
  doi = {10.1523/JNEUROSCI.0517-24.2024},
  urldate = {2026-06-05},
  copyright = {https://creativecommons.org/licenses/by-nc-sa/4.0/},
  langid = {english}
}

@article{thorneVivoDiffusionAnalysis2006,
  title = {{\emph{In Vivo}} Diffusion Analysis with Quantum Dots and Dextrans Predicts the Width of Brain Extracellular Space},
  author = {Thorne, Robert G. and Nicholson, Charles},
  year = 2006,
  month = apr,
  journal = {Proceedings of the National Academy of Sciences},
  volume = {103},
  number = {14},
  pages = {5567--5572},
  issn = {0027-8424, 1091-6490},
  doi = {10.1073/pnas.0509425103},
  urldate = {2025-06-30},
  langid = {english}
}

@article{tonnesenLocalDiffusionExtracellular2023,
  title = {Local Diffusion in the Extracellular Space of the Brain},
  author = {T{\o}nnesen, Jan and Hrab{\u e}tov{\'a}, Sabina and Soria, Federico N.},
  year = 2023,
  month = feb,
  journal = {Neurobiology of Disease},
  volume = {177},
  pages = {105981},
  issn = {09699961},
  doi = {10.1016/j.nbd.2022.105981},
  urldate = {2025-06-26},
  langid = {english}
}

@article{tonnesenSuperResolutionImagingExtracellular2018,
  title = {Super-{{Resolution Imaging}} of the {{Extracellular Space}} in {{Living Brain Tissue}}},
  author = {T{\o}nnesen, Jan and Inavalli, V. V. G. Krishna and N{\"a}gerl, U. Valentin},
  year = 2018,
  month = feb,
  journal = {Cell},
  volume = {172},
  number = {5},
  pages = {1108-1121.e15},
  publisher = {Elsevier},
  issn = {0092-8674, 1097-4172},
  doi = {10.1016/j.cell.2018.02.007},
  urldate = {2025-08-06},
  langid = {english},
  pmid = {29474910}
}

@article{vondiezmannThreeDimensionalLocalizationSingle2017,
  title = {Three-{{Dimensional Localization}} of {{Single Molecules}} for {{Super-Resolution Imaging}} and {{Single-Particle Tracking}}},
  author = {{von Diezmann}, Lexy and Shechtman, Yoav and Moerner, W. E.},
  year = 2017,
  month = jun,
  journal = {Chemical Reviews},
  volume = {117},
  number = {11},
  pages = {7244--7275},
  issn = {0009-2665, 1520-6890},
  doi = {10.1021/acs.chemrev.6b00629},
  urldate = {2022-04-15},
  langid = {english}
}

@article{xieSleepDrivesMetabolite2013,
  title = {Sleep {{Drives Metabolite Clearance}} from the {{Adult Brain}}},
  author = {Xie, Lulu and Kang, Hongyi and Xu, Qiwu and Chen, Michael J. and Liao, Yonghong and Thiyagarajan, Meenakshisundaram and O'Donnell, John and Christensen, Daniel J. and Nicholson, Charles and Iliff, Jeffrey J. and Takano, Takahiro and Deane, Rashid and Nedergaard, Maiken},
  year = 2013,
  month = oct,
  journal = {Science},
  volume = {342},
  number = {6156},
  pages = {373--377},
  publisher = {American Association for the Advancement of Science},
  doi = {10.1126/science.1241224},
  urldate = {2025-07-31}
}

@article{wuConnectingHinderedTransport2020,
  title = {Connecting {{Hindered Transport}} in {{Porous Media}} across {{Length Scales}}: {{From Single-Pore}} to {{Macroscopic}}},
  shorttitle = {Connecting {{Hindered Transport}} in {{Porous Media}} across {{Length Scales}}},
  author = {Wu, Haichao and Wang, Dapeng and Schwartz, Daniel K.},
  year = 2020,
  month = oct,
  journal = {The Journal of Physical Chemistry Letters},
  volume = {11},
  number = {20},
  pages = {8825--8831},
  issn = {1948-7185, 1948-7185},
  doi = {10.1021/acs.jpclett.0c02738},
  urldate = {2026-01-13},
  copyright = {https://doi.org/10.15223/policy-029},
  langid = {english}
}

@article{vanHoveCorrelationsSpaceTime1954,
  author = {Van Hove, L{\'e}on},
  title = {Correlations in Space and Time and Born Approximation Scattering in Systems of Interacting Particles},
  journal = {Physical Review},
  volume = {95},
  pages = {249--262},
  year = {1954},
  doi = {10.1103/PhysRev.95.249}
}

@article{mannTestWhetherOne1947,
  author = {Mann, Henry B. and Whitney, Donald R.},
  title = {On a Test of Whether One of Two Random Variables Is Stochastically Larger than the Other},
  journal = {The Annals of Mathematical Statistics},
  volume = {18},
  number = {1},
  pages = {50--60},
  year = {1947},
  doi = {10.1214/aoms/1177730491}
}

@article{pavaniThreeDimensionalSingleMolecule2009,
  author = {Pavani, Sri Rama P. and Thompson, Michael A. and Biteen, Julie S. and Lord, Samuel J. and Liu, Na and Twieg, Robert J. and Piestun, Rafael and Moerner, W. E.},
  title = {Three-Dimensional, Single-Molecule Fluorescence Imaging beyond the Diffraction Limit by Using a Double-Helix Point Spread Function},
  journal = {Proceedings of the National Academy of Sciences},
  volume = {106},
  number = {9},
  pages = {2995--2999},
  year = {2009},
  doi = {10.1073/pnas.0900245106}
}

@article{dengErgodicPropertiesFractional2009,
  author = {Deng, Weihua and Barkai, Eli},
  title = {Ergodic Properties of Fractional Brownian--Langevin Motion},
  journal = {Physical Review E},
  volume = {79},
  pages = {011112},
  year = {2009},
  doi = {10.1103/PhysRevE.79.011112}
}

@inproceedings{mckinneyDataStructuresStatistical2010,
  author = {McKinney, Wes},
  title = {Data Structures for Statistical Computing in Python},
  booktitle = {Proceedings of the 9th Python in Science Conference},
  pages = {51--56},
  year = {2010},
  doi = {10.25080/Majora-92bf1922-00a}
}

@article{hunterMatplotlib2DGraphics2007,
  author = {Hunter, John D.},
  title = {Matplotlib: A 2D Graphics Environment},
  journal = {Computing in Science \& Engineering},
  volume = {9},
  number = {3},
  pages = {90--95},
  year = {2007},
  doi = {10.1109/MCSE.2007.55}
}

@article{novikovRandomWalksBarriers2014,
  author = {Novikov, Dmitry S. and Fieremans, Els and Jensen, Jens H. and Helpern, Joseph A.},
  title = {Random Walks with Barriers},
  journal = {Nature Physics},
  volume = {10},
  pages = {576--580},
  year = {2014},
  doi = {10.1038/nphys2993}
}

@article{aristovZOLA3DAllowsFlexible2018,
  author = {Aristov, Alexander and Lelandais, Beno{\^i}t and Rensen, Elena and Zimmer, Christophe},
  title = {{ZOLA-3D} Allows Flexible 3D Localization Microscopy over an Adjustable Axial Range},
  journal = {Nature Communications},
  volume = {9},
  pages = {2409},
  year = {2018},
  doi = {10.1038/s41467-018-04709-4}
}

@article{virtanenSciPyFundamentalAlgorithms2020,
  author = {Virtanen, Pauli and others},
  title = {{SciPy} 1.0: Fundamental Algorithms for Scientific Computing in Python},
  journal = {Nature Methods},
  volume = {17},
  pages = {261--272},
  year = {2020},
  doi = {10.1038/s41592-019-0686-2}
}

@article{harrisArrayProgrammingNumPy2020,
  author = {Harris, Charles R. and others},
  title = {Array Programming with {NumPy}},
  journal = {Nature},
  volume = {585},
  pages = {357--362},
  year = {2020},
  doi = {10.1038/s41586-020-2649-2}
}

@misc{allanTrackpy2021,
  author = {Allan, Daniel and Caswell, Thomas and Keim, Nathan and van der Wel, Casper},
  title = {soft-matter/trackpy: trackpy v0.5.0},
  publisher = {Zenodo},
  year = {2021},
  doi = {10.5281/zenodo.4682814}
}

@article{waskomSeabornStatisticalData2021,
  author = {Waskom, Michael L.},
  title = {seaborn: Statistical Data Visualization},
  journal = {Journal of Open Source Software},
  volume = {6},
  number = {60},
  pages = {3021},
  year = {2021},
  doi = {10.21105/joss.03021}
}

\end{document}


\title{Supplementary Information for\\[0.4em]
\textit{Resolving Scale-Dependent Diffusivity in the Brain Extracellular Space}}

\author{Quentin Gresil%
  \authornotes{\textsuperscript{,\textdagger,*}}}
\affiliation{Laboratoire Photonique Num\'erique et Nanosciences,
Universit\'e de Bordeaux, 33400 Talence, France}
\affiliation{LP2N, Institut d'Optique Graduate School,
CNRS UMR 5298, 33400 Talence, France}

\author{Ivo Calaresu}
\affiliation{Interdisciplinary Institute for Neuroscience, CNRS, Univ. Bordeaux, 33076 Bordeaux, France}

\author{Finn L. Sebastian}
\affiliation{Institute for Physical Chemistry, Universität Heidelberg, Heidelberg, Germany}

\author{Benjamin Flavel}
\affiliation{Institute of Nanotechnology, Karlsruhe Institute of Technology, 76131 Karlsruhe, Germany}

\author{Jana Zaumseil}
\affiliation{Institute for Physical Chemistry, Universität Heidelberg, Heidelberg, Germany}

\author{Laurent Groc}
\affiliation{Interdisciplinary Institute for Neuroscience, CNRS UMR 5297, Univ. Bordeaux, 33077 Bordeaux, France}

\author{Laurent Cognet%
  \authornotes{\textsuperscript{,*}}}
\affiliation{Laboratoire Photonique Num\'erique et Nanosciences,
Universit\'e de Bordeaux, 33400 Talence, France}
\affiliation{LP2N, Institut d'Optique Graduate School,
CNRS UMR 5298, 33400 Talence, France}

\begingroup
\renewcommand{\thefootnote}{\fnsymbol{footnote}}
\footnotetext[2]{Present address: Laboratory of Nanoscale Biology (LBEN), Institute of Bioengineering, School of Engineering, Swiss Federal
Institute of Technology Lausanne (EPFL), Lausanne, Switzerland.}
\footnotetext[1]{Correspondence and requests for materials should be
addressed to: Q.G. (\texttt{quentin.gresil@epfl.ch}) or
L.C. (\texttt{laurent.cognet@u-bordeaux.fr}).}
\endgroup

\maketitle

\section{Supplementary Methods}

\subsection{Trajectory representation and time-averaged mean squared displacement}

Three-dimensional trajectories were represented by Cartesian coordinates
$\mathbf{r}(t)=[x(t),y(t),z(t)]$ sampled at
$\Delta t=\SI{0.033}{s}$. For each trajectory, the time-averaged mean squared
displacement (MSD) was calculated as
\begin{equation}
\overline{\delta^2(\tau)}
=
\frac{1}{N_\tau}
\sum_{i=1}^{N_\tau}
\left|
\mathbf{r}(t_i+\tau)-\mathbf{r}(t_i)
\right|^2,
\label{eq:tamsd}
\end{equation}
where $\tau=L\Delta t$ and $N_\tau$ is the number of displacement pairs
available at lag $L$. MSD curves were evaluated for $L=1,\ldots,50$,
corresponding to \SIrange{0.033}{1.650}{s}
\cite{metzlerAnomalousDiffusionModels2014,
rehfeldtRandomWalkersToolbox2023}.

\subsection{Water-calibrated Brownian compatibility test}

For isotropic Brownian diffusion in three dimensions,
$\overline{\delta^2(\tau)}=6D\tau$
\cite{rehfeldtRandomWalkersToolbox2023}. Compatibility with this scaling was
evaluated separately for every trajectory over progressively increasing
fitting windows. For a window containing the first $n$ lag points, the
zero-intercept model
$\overline{\delta^2(\tau_i)}=A_n\tau_i$
was fitted by least squares, with
\begin{equation}
A_n
=
\frac{
\sum_{i=1}^{n}
\tau_i\overline{\delta^2(\tau_i)}
}{
\sum_{i=1}^{n}\tau_i^2
}.
\end{equation}
Departure from linear scaling was quantified by the relative root mean
squared error
\begin{equation}
E_n
=
\left\{
\frac{1}{n}
\sum_{i=1}^{n}
\left[
\frac{
\overline{\delta^2(\tau_i)}-A_n\tau_i
}{
A_n\tau_i
}
\right]^2
\right\}^{1/2}.
\label{eq:relative_error}
\end{equation}

Freely diffusing uCCNTs in water were used to define the experimental null
distribution of $E_n$ separately at every window length. The acceptance
threshold was the empirical 95th percentile of the corresponding water
distribution, using the conservative upper observed quantile. Water
trajectories were evaluated by leave-one-trajectory-out calibration, so that
each trajectory was compared with a threshold obtained from all other valid
water trajectories. Windows containing 2--50 lag points were tested.

Departure was treated as irreversible: once $E_n$ first exceeded its
water-calibrated threshold, the trajectory remained incompatible at all
subsequent windows. This history-dependent rule prevents fluctuations at
poorly sampled long lags from generating repeated transitions between
compatible and incompatible states.

The three categories are operational and relative to the measured window.
A trajectory failing at the earliest tested window was designated
\emph{below-window}; one that failed subsequently was designated
\emph{in-window}; and one that remained compatible through the largest window
was designated \emph{above-window}.

\subsection{Short-time diffusivity and crossover parameters}

The short-time diffusivity was obtained from a zero-intercept fit to the first
three MSD points,
\begin{equation}
D_0=\frac{A_3}{6}.
\end{equation}

For an \emph{in-window} trajectory, the crossover time was the maximum lag of
the first fitting window that irreversibly exceeded the threshold,
\begin{equation}
\tau_c=L_c\Delta t.
\end{equation}
The corresponding exploration length was calculated from the measured MSD,
rather than from a Brownian extrapolation
\cite{wuConnectingHinderedTransport2020},
\begin{equation}
\ell_0
=
\sqrt{
\frac{
\overline{\delta^2(\tau_c)}
}{
3
}
}.
\label{eq:ell0}
\end{equation}
Thus, $\ell_0$ is the root mean squared displacement per Cartesian dimension
at the detected crossover.

Values of $\tau_c$ and $\ell_0$ were not assigned to \emph{below-window} or
\emph{above-window} trajectories. Likewise, an early-lag diffusivity from a
\emph{below-window} trajectory was treated as a resolution-limited effective
value rather than as an unconstrained $D_0$.

\subsection{Ensemble MSD and local scaling exponent}

Time-averaged MSD curves were
averaged at each lag
\cite{metzlerAnomalousDiffusionModels2014,
rehfeldtRandomWalkersToolbox2023} and the local exponent 
\begin{equation}
\alpha(\tau)
=
\frac{
\mathrm{d}
\log
\left\langle
\overline{\delta^2(\tau)}
\right\rangle
}{
\mathrm{d}\log\tau
},
\end{equation}
was evaluated with a centered finite-difference gradient. The complete
trajectory-level analysis was repeated in 500 bootstrap samples; shaded
intervals are the 2.5th and 97.5th percentiles.

\subsection{van Hove distributions and non-Gaussian parameter}

For trajectory $i$, the Cartesian increment at lag $L$ was defined as
\begin{equation}
\delta r_{i,k}(t;L)
=
r_{i,k}(t+L\Delta t)
-
r_{i,k}(t),
\qquad
k\in\{x,y,z\},
\end{equation}
with $\tau=L\Delta t$. Only increments for which both frames were present
were included.

\paragraph{Ensemble-normalized van Hove distribution.}

For the analysis shown in the main text, increments were normalized
coordinate-wise using the lag-dependent root mean squared increment pooled
across all trajectories,
\begin{equation}
\sigma_k(L)
=
\sqrt{
\left\langle
\delta r_{i,k}^{\,2}(t;L)
\right\rangle_{i,t}
}.
\end{equation}
The dimensionless ensemble-normalized displacement was then defined as
\begin{equation}
\Delta x_{i,k}(t;L)
=
\frac{
\delta r_{i,k}(t;L)
}{
\sigma_k(L)
}.
\end{equation}
The normalized increments from all trajectories and the three Cartesian
coordinates were pooled to estimate the ensemble-normalized self-van Hove
distribution,
\begin{equation}
P_{\mathrm{ens}}(\Delta x;L)
=
\left\langle
\delta\!\left[
\Delta x-\Delta x_{i,k}(t;L)
\right]
\right\rangle_{i,t,k}.
\end{equation}

Because the normalization factor was calculated over the complete
population, this distribution retained differences in mobility between
trajectories. More mobile trajectories therefore contributed broader
normalized increments than less mobile trajectories
\cite{vanHoveCorrelationsSpaceTime1954,
wangAnomalousYetBrownian2009}.

\paragraph{Trajectory-normalized van Hove distribution.}

To determine whether the heavy tails arose solely from differences in
mobility between trajectories, a control analysis normalized the increments
of each trajectory separately. For trajectory $i$ and coordinate $k$, the
lag-dependent root mean squared increment was
\begin{equation}
\sigma_{i,k}(L)
=
\sqrt{
\left\langle
\delta r_{i,k}^{\,2}(t;L)
\right\rangle_t
}.
\end{equation}
The trajectory-normalized displacement was defined as
\begin{equation}
\chi_{i,k}(t;L)
=
\frac{
\delta r_{i,k}(t;L)
}{
\sigma_{i,k}(L)
}.
\end{equation}

\begin{figure}[htbp]
\centering
\includegraphics[scale=1]{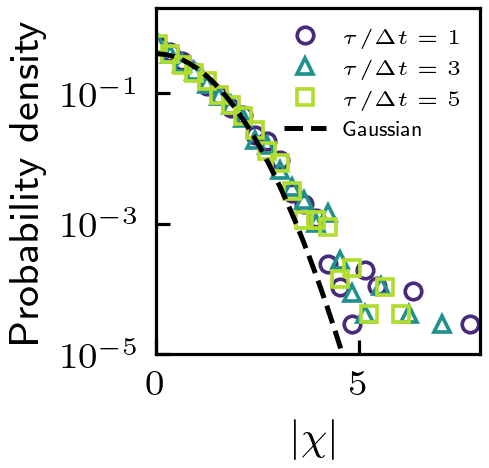}
\caption{\label{fig:trajectory_normalized_vanhove}
\textbf{Trajectory-normalized van Hove distributions of above-window PYR
trajectories.}
Cartesian increments at lag numbers $\tau/\Delta t=1$, 3, and 5 were
normalized separately by the lag-dependent root mean squared increment of
each trajectory before pooling with equal trajectory weight. Only the
nonnegative branch of the signed distribution is displayed. The dashed black
curve denotes the unit-variance Gaussian reference.
}
\end{figure}

For each trajectory, the normalized increments from the $x$, $y$, and $z$
coordinates were pooled to estimate a trajectory-level density
$P_i(\chi;L)$. These densities were then averaged with equal trajectory
weight,
\begin{equation}
P_{\mathrm{tr}}(\chi;L)
=
\frac{1}{N_{\mathrm{tr}}}
\sum_{i=1}^{N_{\mathrm{tr}}}
P_i(\chi;L).
\end{equation}
This normalization removes lag-dependent differences in displacement
variance between trajectories while retaining non-Gaussian displacement
statistics within individual trajectories
\cite{rehfeldtRandomWalkersToolbox2023}. Persistence of heavy tails in
$P_{\mathrm{tr}}(\chi;L)$ therefore indicates that population-level
non-Gaussianity cannot be explained solely by pooling Brownian trajectories
with different constant diffusivities
(Fig.~\ref{fig:trajectory_normalized_vanhove}).

Both distributions were calculated using the complete signed increments.
For visualization, only the nonnegative branch was displayed. The
unit-variance Gaussian and Laplace reference densities were
\begin{equation}
P_{\mathrm G}(q)
=
\frac{1}{\sqrt{2\pi}}
e^{-q^2/2},
\qquad
P_{\mathrm L}(q)
=
\frac{1}{\sqrt{2}}
e^{-\sqrt{2}|q|},
\end{equation}
where $q=\Delta x$ for the ensemble-normalized distribution and $q=\chi$ for
the trajectory-normalized distribution. The Gaussian and Laplace references
were shown for the ensemble-normalized distributions, whereas the
trajectory-normalized control was compared with the Gaussian reference.

\paragraph{Non-Gaussian parameter.}

The departure of the ensemble-normalized increment distribution from a
Gaussian form was quantified using the one-dimensional non-Gaussian
parameter,
\begin{equation}
g(\tau)
=
\frac{
\left\langle
\Delta x^4(\tau)
\right\rangle
}{
3
\left\langle
\Delta x^2(\tau)
\right\rangle^2
}
-1,
\end{equation}
where the moments were calculated after pooling the normalized increments
from the three Cartesian coordinates. A Gaussian displacement distribution
gives $g(\tau)=0$. Positive values indicate a distribution with greater
kurtosis than the Gaussian reference (Fig. S3), arising from a sharper central peak,
heavier tails, or both
\cite{wangAnomalousYetBrownian2009,
rehfeldtRandomWalkersToolbox2023}.

\subsection{Turning angles and increment correlations}

Turning angles were calculated between consecutive non-overlapping
displacements of duration $L\Delta t$,
\begin{equation}
\theta
=
\cos^{-1}
\left(
\frac{
\mathbf v_1\cdot\mathbf v_2
}{
|\mathbf v_1||\mathbf v_2|
}
\right),
\end{equation}
where
$\mathbf v_1=\mathbf r(t+L\Delta t)-\mathbf r(t)$
and
$\mathbf v_2=\mathbf r(t+2L\Delta t)-\mathbf r(t+L\Delta t)$.
Zero-length vectors were excluded. Independent isotropic directions in three
dimensions follow
$P_{\mathrm{iso}}(\theta)=\frac{1}{2}\sin\theta$
\cite{rehfeldtRandomWalkersToolbox2023}.

For increment-correlation analysis, a velocity-like increment was defined as
\begin{equation}
\mathbf u_L(t)
=
\frac{
\mathbf r(t+L\Delta t)-\mathbf r(t)
}{
L\Delta t
}.
\end{equation}
Each trajectory's autocorrelation was normalized by its own zero-separation
variance before averaging. Separation was expressed as $\xi=s/L$, and curves
were fitted over $0\leq\xi\leq4$ with
\begin{equation}
C(\xi;\alpha_C)
=
\frac{1}{2}
\left[
(\xi+1)^{\alpha_C}
+
|\xi-1|^{\alpha_C}
-
2\xi^{\alpha_C}
\right].
\end{equation}
Brownian increments correspond to $\alpha_C=1$, and $\alpha_C<1$ indicates
antipersistence. The fitting form is the normalized increment-correlation
function of fractional Brownian motion
\cite{dengErgodicPropertiesFractional2009}. Fits required at least six valid
points, each receiving contributions from at least 10 trajectories.
Confidence intervals were obtained from 1,000 bootstrap resamples of
complete trajectories.

\subsection{Instantaneous diffusivity and post-crossover restriction}

For the \emph{in-window} category, instantaneous diffusivity was defined from
the ensemble MSD as
\begin{equation}
D_{\mathrm{inst}}(\tau)
=
\frac{1}{6}
\frac{
\mathrm{d}
\left\langle
\overline{\delta^2(\tau)}
\right\rangle
}{
\mathrm{d}\tau
}.
\end{equation}
The derivative was evaluated with a second-order finite-difference gradient;
only finite positive values were used for logarithmic fitting. The
derivative-based definition makes the dependence of apparent diffusivity on
observation time explicit
\cite{novikovRandomWalksBarriers2014}.

The post-crossover decrease was summarized by
\begin{equation}
D_{\mathrm{inst}}(\tau)
=
A\tau^{-\vartheta}.
\label{eq:vartheta}
\end{equation}
The regression used lag numbers 7--40, corresponding to
\SIrange{0.231}{1.320}{s}. This interval was selected from candidate final
lags 20, 25, 30, 35, 40, 45, and 50 by considering log-space fit quality,
fit coverage, and stability of the regional contrast.

\subsection{Statistical analysis and resampling}

Regional inference used tissue slices, rather than trajectories, as the
summarized biological units. Parameters were summarized by their median in
each condition--category--slice combination, and combinations containing
strictly more than 10 valid trajectories were eligible. PYR and RAD were
compared using two-sided Mann--Whitney $U$ tests
\cite{mannTestWhetherOne1947}.

Uncertainty in the pooled $\vartheta$ estimate was assessed using 1,000
hierarchical slice-bootstrap samples. Eligible slices were sampled with
replacement within each condition, retaining all trajectories in each
selected slice. The ensemble MSD, $D_{\mathrm{inst}}$, and $\vartheta$ were
recomputed for every replicate. Percentile confidence intervals correspond
to the 2.5th and 97.5th percentiles.

The regional contrast was evaluated in each joint replicate as
\begin{equation}
\Delta\vartheta
=
\vartheta_{\mathrm{PYR}}
-
\vartheta_{\mathrm{RAD}}.
\end{equation}
A total of 11 slices from 3 animals were used such that biological statistics are made at the slice level.

Analyses were performed in Python
\cite{harrisArrayProgrammingNumPy2020,
mckinneyDataStructuresStatistical2010,
virtanenSciPyFundamentalAlgorithms2020,
allanTrackpy2021,
hunterMatplotlib2DGraphics2007,
waskomSeabornStatisticalData2021}.

\section{Supplementary Validation and Sensitivity Analyses}

\subsection{Water calibration of the progressive Brownian test}

\begin{figure}[htbp]
\centering
\includegraphics[scale=1]{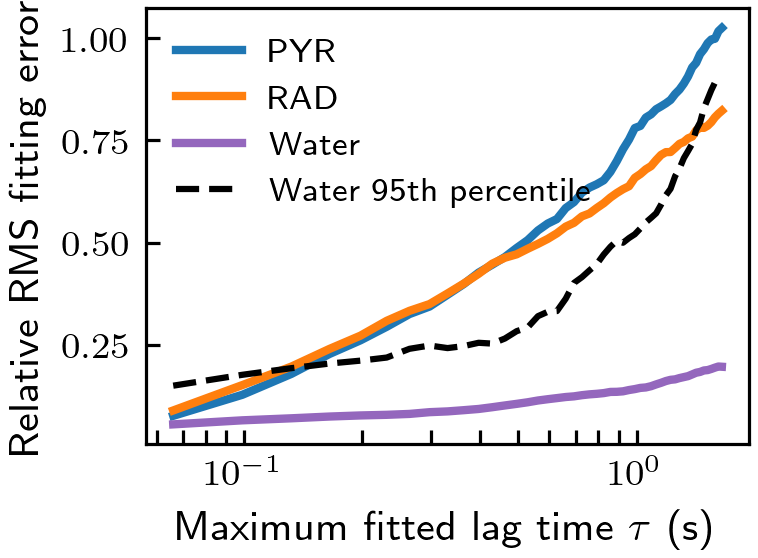}
\caption{\label{fig:classifier_validation}
\textbf{Water calibration of the progressive Brownian compatibility test.}
Median relative RMS fitting error for PYR, RAD, and leave-one-out water
trajectories as the maximum fitted lag time was extended. The dashed black
curve shows the empirical 95th percentile of the water error distribution
used as the window-specific acceptance threshold.
}
\end{figure}

The relative RMS fitting error increased as additional lag points were
included in the Brownian fit. The empirical 95th percentile of the water
error distribution was therefore calculated separately at every fitting
window and used as the corresponding acceptance threshold
(Fig.~\ref{fig:classifier_validation}). This window-specific calibration
accounts for the finite-sample curvature and measurement noise present in
the experimental water control.

\subsection{Selection of the transport-hindrance fitting interval}

The transport-hindrance exponent, $\vartheta$, was estimated over a common
lag interval of $L=7$--40, corresponding to
$\tau=\SIrange{0.231}{1.320}{s}$. The starting lag was fixed at $L=7$ to
exclude the crossover transition and restrict the fit to the subsequent
decline in instantaneous diffusivity. The same starting and final lags were
used for every slice and condition, preventing differences in fitted
timescale from contributing to the regional comparison.

\begin{figure}[htbp]
\centering
\includegraphics[scale=1]{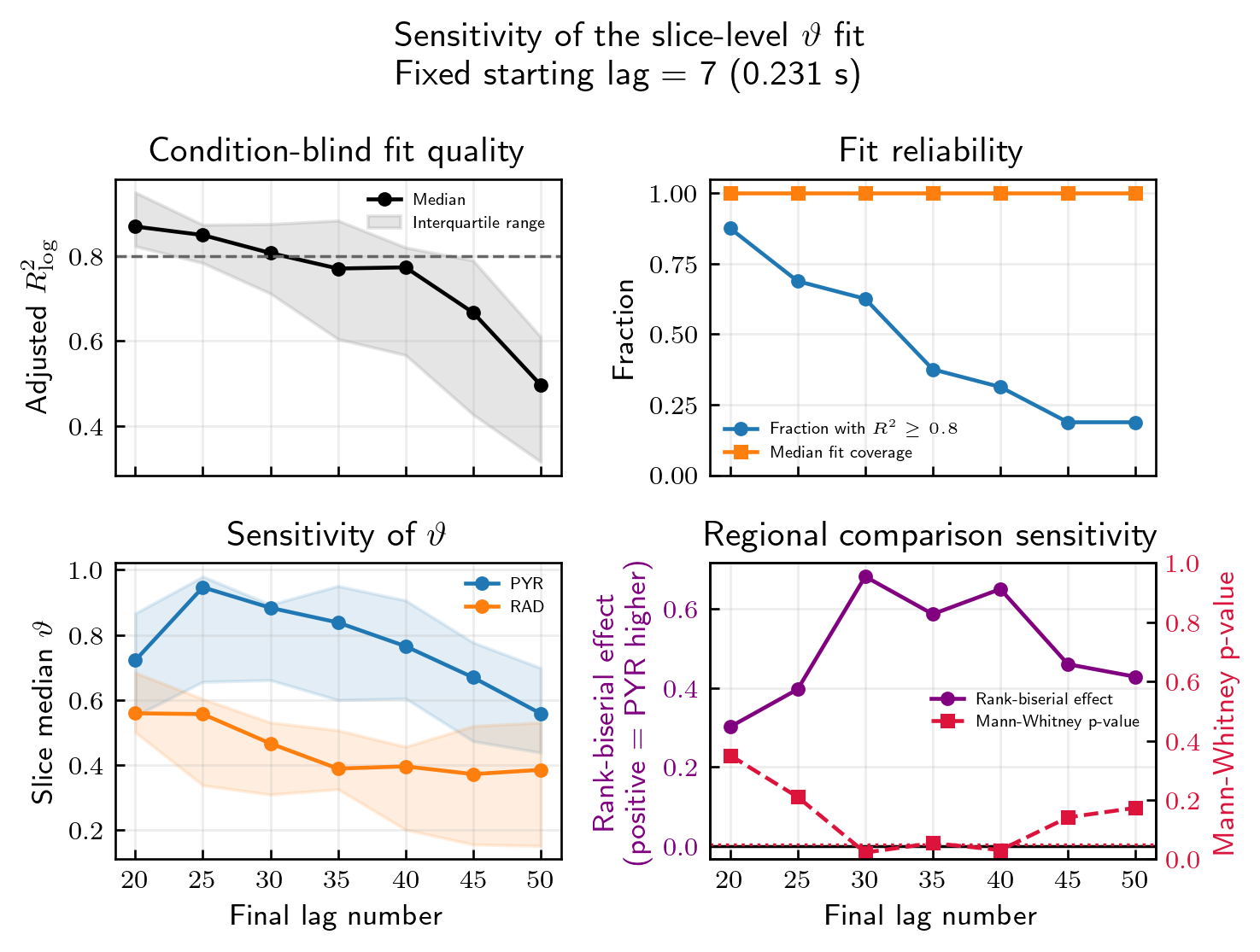}
\caption{\label{fig:vartheta_sensitivity}
\textbf{Sensitivity of the post-crossover restriction exponent,
$\vartheta$, to the fitting interval.}
The starting lag was fixed at $L=7$ ($\tau=0.231~\mathrm{s}$), while the final
fitted lag was varied from $L=20$ to $L=50$.
(a) Median adjusted log-space coefficient of determination,
$R_{\log}^{2}$, across eligible slice-level fits. Shading denotes the
interquartile range, and the dashed line marks $R_{\log}^{2}=0.8$.
(b) Fraction of slice-level fits with $R_{\log}^{2}\geq0.8$ and median
fraction of the requested fitting interval represented by valid data.
(c) Slice-level median $\vartheta$ in PYR and RAD as a function of the final
fitted lag. Shaded regions denote the interquartile ranges across slices.
(d) PYR--RAD comparison as a function of the final fitted lag. The purple
curve shows the rank-biserial effect size, defined as positive when
$\vartheta_{\mathrm{PYR}}>\vartheta_{\mathrm{RAD}}$, and the red dashed curve
shows the two-sided Mann--Whitney $P$ value.
}
\end{figure}

The final lag was selected using condition-blind fit diagnostics rather than
the regional effect size or statistical significance
(Fig.~\ref{fig:vartheta_sensitivity}). Shorter intervals produced higher
adjusted coefficients of determination but sampled a more restricted range
of the post-crossover evolution. Estimates of $\vartheta$ and their regional
ordering were comparatively stable when the final lag was varied between
$L=30$ and $L=40$. Extending the fit beyond $L=40$ produced a marked loss of
log--log fit quality and a reduction in the fraction of slices with
$R_{\log}^{2}\geq0.8$. We therefore used $L=40$ as the longest candidate
endpoint preceding this deterioration.

The fitted power law should consequently be interpreted as a compact summary
of the average decline in $D_{\mathrm{inst}}(\tau)$ over this fixed interval,
rather than as evidence that every individual slice follows an exact power
law over the complete experimental range.

\clearpage

\section{Supplementary Figures}

\begin{figure}[htbp]
\centering
\includegraphics[scale=1]{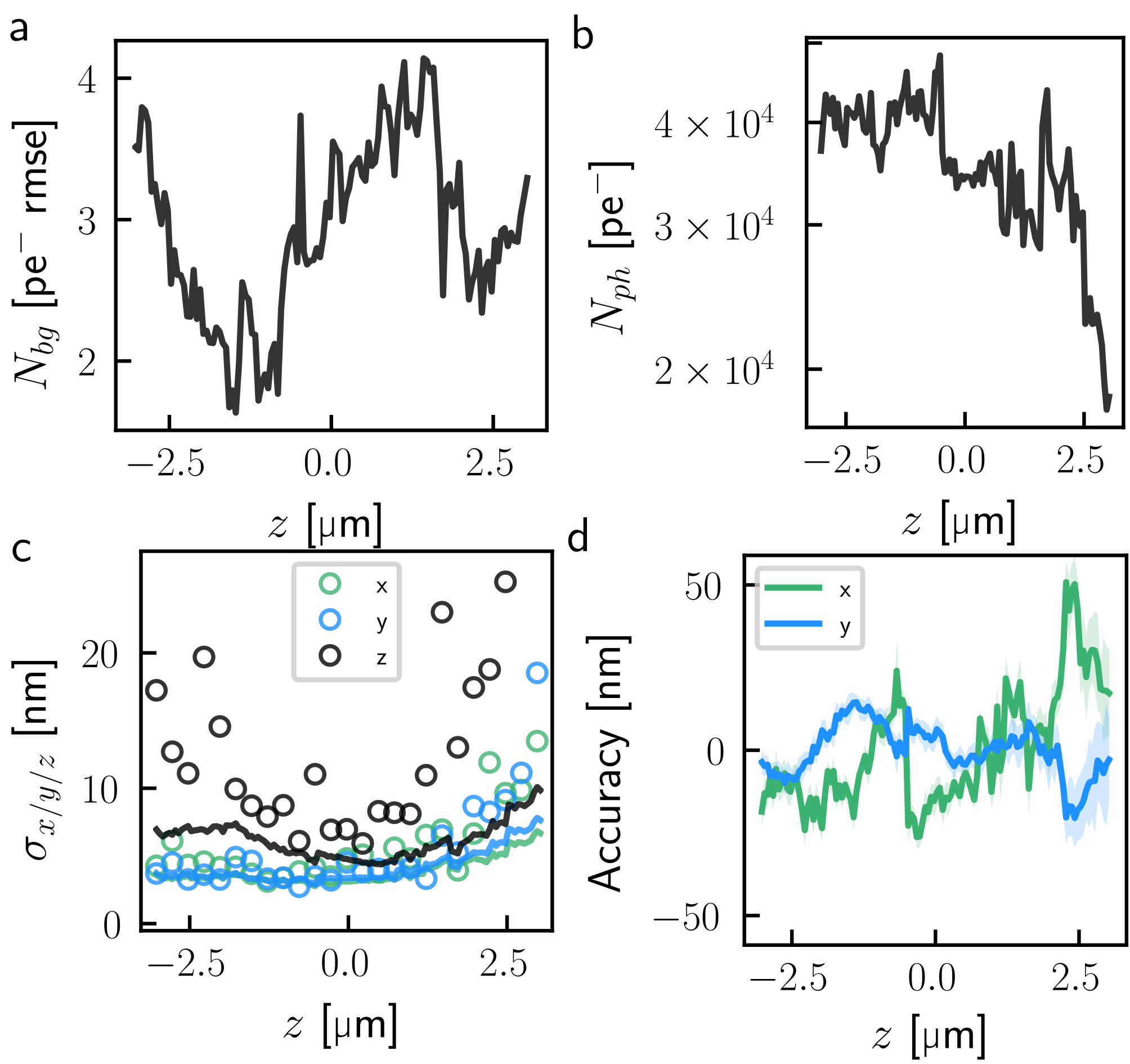}
\caption{\label{fig:localization_precision}
\textbf{Experimental calibration and localization performance of the
double-helix point-spread function.}
An immobilized uCCNT was imaged over an axial range of approximately
$\pm 3~\si{\micro\metre}$, with 50 repeated acquisitions at each axial
position.
(a,b) Fitted background level, $N_{\mathrm{bg}}$, and detected photon number,
$N_{\mathrm{ph}}$, respectively, as functions of the imposed axial position.
(c) Experimental localization precision in $x$, $y$, and $z$, calculated as
the standard deviation of the repeated localizations at each axial position
(open circles). Solid lines show the corresponding Cramér--Rao lower bounds
calculated from the retrieved PSF, fitted photon number, and background
level.
(d) Lateral localization accuracy, expressed as the displacement of the mean
localized $x$ and $y$ positions from their reference positions as the emitter
was translated axially.
}
\end{figure}

Double-helix point-spread-function calibration and three-dimensional
localization followed established single-molecule localization procedures
\cite{pavaniThreeDimensionalSingleMolecule2009,
vondiezmannThreeDimensionalLocalizationSingle2017,
aristovZOLA3DAllowsFlexible2018}
(Fig.~\ref{fig:localization_precision}).

\clearpage

\section{Supplementary Tables}

\begin{table}[htbp]
\centering
\caption{\label{tab:dataset_summary}
\textbf{Summary of the final datasets and operational categories.}}
\begin{tabular}{lrrrr}
\toprule
Condition & Total & Below-window & In-window & Above-window \\
\midrule
PYR   & 721 & 246 (34.1\%) & 335 (46.5\%) & 140 (19.4\%) \\
RAD   & 958 & 354 (37.0\%) & 392 (40.9\%) & 212 (22.1\%) \\
Water & 591 & -- & -- & -- \\
\bottomrule
\end{tabular}
\end{table}

\begin{table}[htbp]
\centering
\caption{\label{tab:slice_inclusion}
\textbf{Cohorts used for slice-level regional inference.}
Eligibility required strictly more than 10 trajectories in the corresponding
slice and category. The slightly smaller $\vartheta$ cohort reflects the
additional requirement of a valid post-crossover instantaneous-diffusivity
fit. Counts are based on the stored analysis tables.
}
\begin{tabular}{lllrr}
\toprule
Condition & Category & Analysis & Eligible slices & Trajectories \\
\midrule
PYR & In-window    & $D_0$, $\ell_0$ & 7 & 322 \\
RAD & In-window    & $D_0$, $\ell_0$ & 9 & 383 \\
PYR & In-window    & $\vartheta$      & 7 & 319 \\
RAD & In-window    & $\vartheta$      & 9 & 381 \\
PYR & Above-window & $D_0$            & 5 & 129 \\
RAD & Above-window & $D_0$            & 6 & 177 \\
\bottomrule
\end{tabular}
\end{table}

\bibliographystyle{apsrev4-2}
\bibliography{supplementary_references}